\documentclass[%
reprint,
amsmath,amssymb,
aps,
]{revtex4-2}

\usepackage{graphicx}
\usepackage{dcolumn}
\usepackage{bm}
\usepackage[dvipsnames]{xcolor}
\usepackage[normalem]{ulem}
\usepackage{hyperref}

\usepackage{subfig}

\usepackage{cancel}
\usepackage{soul}

\begin{document}

	\setstcolor{red}
	\preprint{APS/123-QED}
	
	\title{Light rings on stationary axisymmetric spacetimes:\\blind to the topology and able to coexist}
	
	\author{Pedro V. P. Cunha}
	\author{Carlos A. R. Herdeiro}
	\author{João P. A. Novo}%
	\affiliation{%
		Departamento de Matem\'atica da Universidade de Aveiro \\
		and Centre for Research and Development  in Mathematics and Applications (CIDMA),\\ Campus de Santiago, 3810-183 Aveiro, Portugal
	}%

	\date{\today}
	\begin{abstract}
		It has been established that Black Hole (BH) spacetimes obeying some general set of assumptions always possess, at least, one light ring (LR), per rotation sense~\cite{Cunha:2020azh}. This theorem was originally established  for asymptotically flat, stationary, axial symmetric, 1+3 dimensional circular spacetimes harbouring a non-extremal and topologically spherical Killing horizon. Following the mantra that a theorem is only as strong as its assumptions in this work we extend this theorem to non topologically spherical (toroidal) BHs and to spacetimes harbouring more than one BH. As in~\cite{Cunha:2020azh}, we show that each BH still contributes with, at least, one LR (per rotation sense).
	\end{abstract}
	
	\maketitle
	
	\tableofcontents{}
	
	\section{Introduction}
	
	The shadow and ringdown of a Black Hole (BH) are closely connected to its light rings (LRs)~\cite{Cardoso:2016rao,Cunha:2018acu}. Hence, in this new era of precision gravity where it is possible to directly measure gravitational wave events corresponding to the collision of BHs, as well as directly seeing the shadows of supermassive BHs, it is essential to deepen our understanding of LRs.

    It has been established that BH spacetimes must contain at least one LR (per rotation sense), provided that some general set of assumptions are satisfied~\cite{Cunha:2020azh} (see also~\cite{hod2013upper,paganini2018role}).
    When such assumptions are violated, it might be possible to have non-trivial results. For instance, it has been established that the asymptotic structure of the spacetime has implications on the number of LRs around BHs~\cite{Junior:2021svb}. In this regard this paper aims to shed some light on the interplay between the BH horizon's topology and its LRs, \emph{i.e.} to inquire whether a BH with a non spherical topology obeys the same theorem as a spherical one. Moreover, it is interesting to assess the robustness of such existence results of LRs when several BHs coexist in the same spacetime. Physically, one expects that when two BHs are very far apart they will essentially be unaffected by each other, hence each will have its own LR. But is this expectation true given the non-linearity of the theory? If so, does the separation distance play a role? These are questions we aim to address in the present work.
	
	The topology of event horizons has been the focus of several theorems in the literature. For instance, Hawking showed in a seminal work that  any stationary event horizon is topologically a sphere and if it is spinning it must be axisymmetric, provided that certain physically reasonable assumptions hold ($e.g.$ Dominant Energy Condition)~\cite{Hawking:1971vc,hawking_ellis_2023}. Later Gannon was able to drop the assumption of stationarity and showed that, when the spacetime is allowed to be dynamical, event horizons can be either spherical or toroidal~\cite{Gannon1976}. However, any putative such toroidal structures should be short lived according to the topological censorship theorem~\cite{Friedman:1993ty}: if valid (see~\cite{Krasnikov:2010vw}) then no observer can probe the topology of the spacetime, hence the "hole" of the torus would have to collapse before any causal curve would be able to cross it. This theorem also makes assumptions on the matter content of the spacetime, namely the Null Energy condition must be satisfied. In fact, it has been predicted that for generic BH binary systems the event horizon is very briefly toroidal  at some point in the evolution of the system~\cite{PhysRevD.58.104016}. Such short lived toroidal event horizons have indeed been observed in binary BH merger simulations~\cite{Bohn:2016soe,Husa_1999}. 
 
 Some findings in this paper are concerned with the properties of stationary toroidal BHs. Considering the results mentioned above, it becomes apparent that these toroidal BHs inherently challenge several energy conditions with respect to the effective energy-momentum tensor. Nevertheless, since the theorem presented in this paper does not rely on specific field equations, its validity extends across any metric theory of gravity, thereby leading us to overlook these violations.

	General relativity solutions containing multiple BHs have been known for quite some time. In fact, Bach and Weyl proposed in 1922 a solution describing two Schwarzschild BHs placed at some finite distance in 4D asymptotically flat spacetime~\cite{BachWeyl}. This solution however is plagued with a conical singularity preventing the collapse of the two event horizons. This was latter generalised to solutions containing $N$ collinear neutral static BHs by Israel and Kahn~\cite{1964NCim...33..331I}. This solution still requires conical singularities,  except in the $N\rightarrow\infty$ limit, which can be interpreted as a higher dimensional BH in a compactified spacetime~\cite{PhysRevD.35.455}. 
 
 By introducing spin it is possible to construct a double-Kerr vacuum solution by resorting to solution generation techniques, like inverse scattering~\cite{KRAMER1980259}. When the BHs are co-rotating, \emph{i.e} the spins are aligned, the spin-spin interaction is repulsive~\cite{Wald:1972sz,Herdeiro:2008kq}, but this is not enough to maintain an equilibrium between the BHs without resorting to conical singularities~\cite{Costa:2009wj,Hennig:2019knn}. 
  Another way to construct multiple BH solutions is to immerse them in an external gravitational field~\cite{Vigano:2022hrg,Astorino:2021dju,Annulli:2023ydz}, but asymptotic flatness is lost in those cases. 
	
	If one considers electrovacuum instead there is a class of solutions that evade all these problems, the Majumdar-Papapetrou solutions~\cite{Majumdar:1947eu,Papapetrou:1948jw}. These represent an arbitrary number of charged BHs in equilibrium in an asymptotically flat 4D spacetime regular on and outside the event horizons. In this case is the electric charge of the BHs that counteracts the gravitational attraction and keeps them in equilibrium. However, all the BHs must be extremal.

	Recently, inspired by BH solutions with scalar hair~\cite{Herdeiro:2014goa}, it was found that scalar fields can in fact balance the gravitational attraction yielding two BH asymptotically flat solutions without the need for conical singularities or extremal BHs. This was achieved for both static~\cite{Herdeiro:2023mptSch} and for spinning BHs~\cite{Herdeiro:2023roz}. Only for the latter case, however, the scalar field obeys the weak energy condition.
	
	The main goal of this work is to establish a theorem on the existence and number of LRs on a stationary and axial symmetric spacetime containing an arbitrary number of non-extremal event horizons. The paper is structured as follows: in Sec. \ref{sec:space} the assumptions on the spacetime are introduced and discussed, and a general form for the metric is presented. In Sec. \ref{sec:Form} the formalism of identifying the LRs of a spacetime as critical points of a potential defined on the 2-plane orthogonal to the Killing plane is discussed. After asserting the formalism and the assumptions of this paper we will present each generalization of the initial theorem~\cite{Cunha:2020azh} in a different section. In Sec. \ref{se:SingTor} a generic spacetime with single toroidal BH is considered, with an example of such a metric presented at the end to illustrate our result. In Sec. \ref{sec:MultTor} the case of multiple toroidal BHs is discussed. Sec. \ref{sec:MultiBHs} is devoted to discussing the construction used to generate spacetimes with multiple BHs and the number of LRs on such spacetimes. All is then put together in Sec. \ref{sec:Everything} where spacetimes containing an arbitrary number of toroidal and spherical BHs are considered.

	\section{The spacetime\label{sec:space}}
	
	We consider spacetimes which are asymptotically flat, stationary and axisymmetric, meaning that they possess two Killing vector fields: a time-like vector at infinity, $\xi$, and a space-like vector, $\eta$, associated with time translations and rotations, respectively. In addition, the spacetimes is assumed to contain one (or more) BH Killing horizons. As was shown by Carter, if the spacetime is asymptotically flat the vector fields $\{\xi, \eta\}$ commute~\cite{Carter:1970ea}, and it is possible to define a coordinate system adapted to both the symmetries simultaneously, meaning that $\eta=\partial_t$ and $\xi=\partial_\phi$. Further imposing circularity of the spacetime its line element can be written as, \emph{e.g.} sec. 7.1 of~\cite{Wald:1984rg}:	
	\begin{equation}
		{\rm d}s^{2}=g_{tt}{\rm d}t^{2}+2g_{t\phi}{\rm d}\phi{\rm d}t+g_{\phi\phi}{\rm d}\phi^{2}+g_{\rho\rho}{\rm d}\rho^{2}+g_{zz}{\rm d}z^{2}\,.\label{eq:LineEl}
	\end{equation}
	Where $\left\{\rho,z\right\}$ are the canonical Bach-Weyl coordinates, which reduce to the usual cylindrical ones at spatial infinity.
	
	The rotation axis is defined as the set of points which is left invariant by the action of the angular Killing vector field, $\eta$, hence at the axis 
	\begin{equation}
		\begin{cases}
			g_{\phi\phi}=\eta\cdot\eta=0\,,\\
			g_{t\phi}=\eta\cdot\xi=0\,.
		\end{cases}\label{eq:RotAxis}
	\end{equation}
	The radial coordinate $\rho$ is defined such that the rotation axis is located at $\rho=0$.
	
	\section{Topological charge of LRs\label{sec:Form}}
	
	In this paper a LR is defined as a null geodesic whose tangent vector, $k$, is a linear combination of the time translation and axial Killing vectors.
	It was shown previously~\cite{Cunha:2017qtt} that these orbits can be associated with the critical points of two effective potentials (one per rotation sense of the horizon), $H_\pm$, on the orthogonal subspace spanned by $\rho$ and $z$, given by
	\begin{equation}
		H_\pm\left(\rho,z\right):=\frac{-g_{t\phi}\pm\sqrt{D}}{g_{\phi\phi}}\,,\quad D:=g_{t\phi}^2-g_{tt}g_{\phi\phi}\label{eq:2dPot} \ .
	\end{equation}
	
	This equivalence can be used to assign a topological charge to each LR. To do so one considers the normalized gradient of $H_\pm$ given by:
	\begin{equation}
		\mathbf{v}=\left(v_\rho,v_z\right)=\left(\frac{\partial_\rho H_\pm}{\sqrt{g_{\rho\rho}}},\frac{\partial_z H_\pm}{\sqrt{g_{zz}}}\right)\,. \label{eq:VecField}
	\end{equation}
	The normalization is such that $\partial_\mu H_\pm\partial^\mu H^\pm=v_\rho^2+v_z^2:=v^2$.	At the critical points of the effective potentials one has $\mathbf{v}=0=v$, so this is the condition for the existence of a LR. The definition of a topological charge requires the construction of a contour, $\mathcal{C}$, on the $\{\rho,z\}$ plane, which is simple, closed and piece-wise smooth. In addition, we can consider an auxiliary bi-dimensional Cartesian space $\{v_\rho,v_z\}$, denoted $\mathcal{V}$. When traversing $\mathcal{C}$ in the positive sense a curve, $\tilde{\mathcal{C}}$, will be defined in $\mathcal{V}$. Since $\mathcal{C}$ is closed $\tilde{\mathcal{C}}$ will also be closed. Introducing polar coordinates on $\mathcal{V}$ as $v_\rho=v\cos\Omega\,,v_z=v\sin\Omega$, the winding number $w$ of $\mathbf{v}$ is defined as the total variation of $\Omega$ divided by $2\pi$, as $\mathcal{C}$ is circulated in the positive sense:
	\begin{equation}
		w=\frac{1}{2\pi}\oint_\mathcal{C}\mathrm{d}\Omega\,.
	\end{equation}
	Non-degenerate critical points \footnote{Here a distinction is made between degenerate and non-degenerate critical points. A critical point is said to be degenerate if the determinant of the Hessian vanishes at that point. In this work we neglect the analysis of such points since the occurrence of such points requires very specific conditions, and make the analysis more nuanced. For example, the existence of such a point may allow an ultracompact object to possess a single LR \cite{Hod:2017zpi}, corresponding to the coalescence of the two LRs predicted by \cite{Cunha:2017qtt}} of $H_\pm$ have $w=\pm 1$, where $w=1$ corresponds to a maximum or minimum and $w=-1$ corresponds to a saddle point. If a contour encircles several critical points, then the total topological charge is the sum of the individual topological charges associated with each one of the critical points.
	
	The identification of LRs as critical points of the potentials $H_\pm$ has been used  to prove existence results for LRs, and their stability, for different kinds of spacetimes, with or without BHs~\cite{Cunha:2018acu,Cunha:2017qtt,Junior:2021svb}. In \cite{Ghosh:2021txu} instead of focusing on the effect of an event horizon an ergosurface was considered, and it was shown that if a spacetime satisfying similar assumptions as described above possesses an ergosurface it will have at least one LR outside the ergosurface. There have also been generalizations to higher dimensional spacetimes \cite{Tavlayan:2022hzl}. More recently, similar arguments have been extended to the study of timelike circular orbits~\cite{Wei:2022mzv, Yin:2023pao}. In a different direction, this approach has also inspired a new field of study where BHs are treated as defects in the thermodynamical parameter space, wherein a topological charge can be associated to each defect, \emph{i.e.} BH solution~\cite{Wei:2022dzw}.

	\section{Toroidal BH\label{se:SingTor}}
	\subsection{The Killing horizon}
	
	The Killing horizon, $\mathcal{H}$, corresponds to the translation along the Killing vector fields of a regular closed curve, $\mathcal{H}^*$ defined on the $\{\rho,z\}$ plane. In order to have a well defined topology, this curve should be simple (\emph{i.e} it cannot have self-intersections), and no point in the curve should lie on the rotation axis. 
	
	Having a BH Killing horizon means there is a Killing vector field, $\chi=\xi+\omega_H \eta,\:\omega_H=\mathrm{const.}$, which is null on $\mathcal{H}$, \emph{i.e.} $\left.\chi^\mu\chi_\mu\right|_\mathcal{H}=0$. In particular $\chi_{\mu}\chi^{\mu}$ is constant on the horizon. This means that $\nabla^{\alpha}\left(\chi_{\mu}\chi^{\mu}\right)$ is also normal to the horizon, hence, there must exist some $\kappa$ such that $\left.\nabla^{\alpha}\left(\chi_{\mu}\chi^{\mu}\right)\right|_\mathcal{H}=-2\kappa \left.\chi^{\alpha}\right|_\mathcal{H}$ at the horizon. The value of $\kappa$ is constant over both orbits of $\chi$ and also over the horizon, and is known as the \emph{surface gravity} of $\mathcal{H}$~\cite{Wald:1984rg}. Since $\chi_\mu\chi^\mu$ does not depend on $t$ and $\phi$, one has
	\begin{equation}
		\begin{cases}
			0&=\left.\left(g_{\phi t}+\omega_H g_{\phi\phi}\right)\right|_\mathcal{H}\,,\\
			0&=\left.\left( g_{tt}+\omega_H g_{t\phi}\right)\right|_\mathcal{H}\,.
		\end{cases}
	\end{equation}
	From where it is possible to deduce $\omega_H=\left.-g_{t\phi}/g_{\phi\phi}\right|_{\mathcal{H}}$ and $\left.D\right|_{\mathcal{H}}=\left.\left(g_{t\phi}^2-g_{tt} g_{\phi\phi}\right)\right|_\mathcal{H}\equiv0$. Since $D$ corresponds to minus the determinant of the Killing part of the metric, and provided the metric outside $\mathcal{H}$ has a Lorentzian signature, then one has $D>0$ outside the horizon.
	
	\subsection{The contour}

	
	The toroidal topology of the horizon implies that, when working in cylindrical coordinates, it is possible to define a contour on the non-Killing plane, denoted $\mathcal{C}$, such that $\mathcal{H}^*$ lies on its interior, denoted $\mathcal{E}$. In the appropriate limit $\mathcal{E}$ will correspond to the entire orthogonal 2-space. The contour will be defined as the union of 4 line segments, $\mathcal{I}_{i=1,2,3,4}$, defined as $\mathcal{I}_1=\{\rho=\delta,-h<z<h\},\,\mathcal{I}_2=\{\delta<\rho <R,z=-h\},\,\mathcal{I}_3=\{\rho=R,-h<z<h\},\,\mathcal{I}_4=\{\delta<\rho< R,z=h\} $. This construction is illustrated in Fig. (\ref{fig:IntCon}). 
	
	We are only interested in the spacetime outside $\mathcal{H}$, the contributions from its interior must then be removed from the total topological charge. To do so one begins by deforming $\mathcal{C}$ until it matches $\mathcal{H}^*$, this allows for the computation of the winding number along that deformed contour, denoted $w_\mathcal{H}$. Finally, the topological charge outside the event horizon, $w_\mathcal{I}$, will be the total charge of the entire spacetime with the contribution from the horizon subtracted, \emph{i.e.} $w_\mathcal{E}=w_{I_1}+w_{I_2}+w_{I_3}+w_{I_4}-w_\mathcal{H}$, where
	\begin{align}
		w_{I_1}&=\frac{1}{2\pi}\intop_{h}^{-h} \left.\frac{\mathrm{d}\Omega}{\mathrm{d}z}\right|_{\rho=\delta}\mathrm{d}z\,,\quad w_{I_2}=\frac{1}{2\pi}\intop_{\delta}^R \left.\frac{\mathrm{d}\Omega}{\mathrm{d}\rho}\right|_{z=-h}\mathrm{d}\rho\,,\nonumber\\
		w_{I_3}&=\frac{1}{2\pi}\intop_{-h}^h \left.\frac{\mathrm{d}\Omega}{\mathrm{d}z}\right|_{\rho=R}\mathrm{d}z\,,\quad w_{I_4}=\frac{1}{2\pi}\intop_R^{\delta} \left.\frac{\mathrm{d}\Omega}{\mathrm{d}\rho}\right|_{z=h}\mathrm{d}\rho   \,, \\
		w_\mathcal{H}&=\frac{1}{2\pi}\oint_\mathcal{H}\mathrm{d}\Omega\nonumber
	\end{align}
	Then the total topological charge outside the BH is obtained when taking the limits, $h\rightarrow+\infty,\delta\rightarrow0,R\rightarrow+\infty$.
	
	The behaviour of the vector field must be studied in three different limits, the asymptotic ($\mathcal{I}_{2,3,4}$), the axis ($\mathcal{I}_{1}$) and horizon. Each of these will be addressed below.

	\begin{center}
		\begin{figure}
			\begin{centering}
				\includegraphics[width=0.8\linewidth]{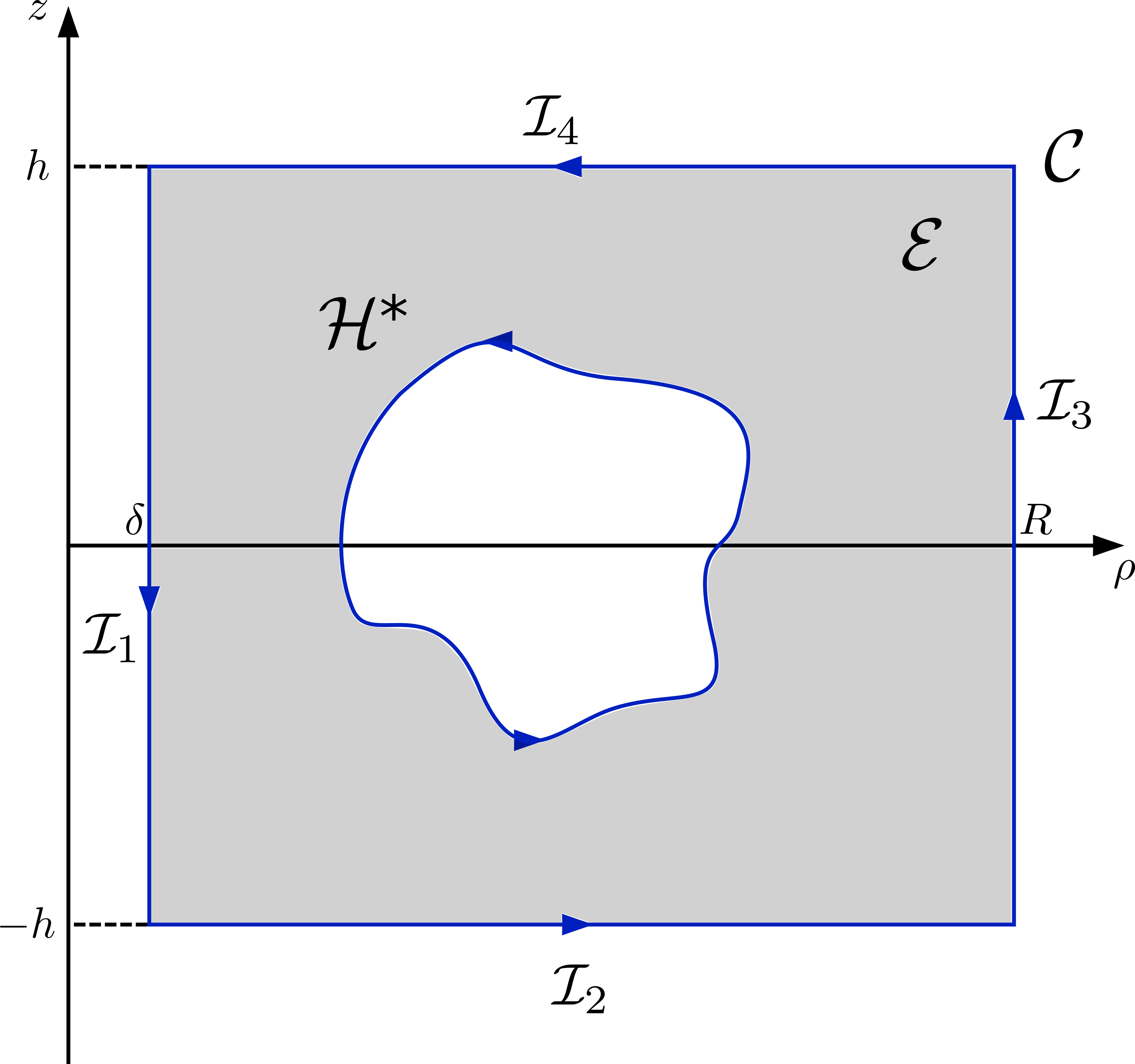}
				\par\end{centering}
			\caption{Illustration of the integration contour, $\mathcal{C}$, corresponding to the union of the $I_i$, defined in the $\left\{\rho,z\right\}$ plane. The interior of the contour, $\mathcal{E}$, as well as the cross section of the event horizon $\mathcal{H}^*$, is defined on $t=\mathrm{const.},\,\phi=\mathrm{const.}$ hypersurfaces.\label{fig:IntCon}}
		\end{figure}
		\par\end{center}

	\subsubsection{Asymptotic limit}
	
	In cylindrical coordinates spatial infinity occurs when either
	\begin{equation}
		\rho\rightarrow\infty\,,z\rightarrow\pm\infty\,.
	\end{equation}
	Where Minkowski spacetime is recovered, in cylindrical coordinates this corresponds to
	\begin{equation}
		{\rm d}s{{}^2}=-{\rm d}t^{2}+\rho^{2}{\rm d}\phi^{2}+{\rm d}\rho^{2}+{\rm d}z^{2}\,.
	\end{equation}
	Therefore, the effective potentials behave in this limit as
	\begin{equation}
		H_\pm\simeq\pm\frac{1}{\rho}\,,
	\end{equation} 
	and the vector fields ${\bf v}^\pm$ go like 
	\begin{equation}
		\begin{cases}
			v_{\rho}^{\pm}\rightarrow\mp\frac{1}{\rho^{2}}\Rightarrow\frac{v_{\rho}^{\pm}}{\left|\mathbf{v}^{\pm}\right|}\rightarrow\mp1\,,\\
			v_{z}^{\pm}\rightarrow0\,.
		\end{cases}
	\end{equation}
	Therefore, when taking $h,R\rightarrow\infty$, $\mathbf{v}$ will be constant along $\mathcal{I}_2,\,\mathcal{I}_3$ and $\mathcal{I}_4$, thus $w_{I_2,I_3,I_4}=0$.

	\subsubsection{Axis limit}

	The definition of the rotation axis, Eq. (\ref{eq:RotAxis}), states that $g_{\phi\phi}=0=g_{t\phi}$. How these functions approach zero can be constrained if one imposes some degree of regularity at the axis. The absence of conical singularities on the axis implies that near the axis 
	\footnote{
		The absence of conical singularities corresponds to requiring that sufficiently near the axis the proper perimeter, $\mathcal{P}$, and radius, $\mathcal{R}$, of a small circumference of constant $t,r,z$ are related by the  usual Euclidean formula, $\mathcal{P}=2\pi\mathcal{R}$. For the line element (\ref{eq:LineEl}) we have
		\begin{equation}
			\mathcal{P} 
			= 2\pi\sqrt{g_{\phi\phi}}\,,\; 		\mathcal{R} = \intop_0^\rho\sqrt{g_{\rho\rho}}\mathrm{d}\rho \simeq \rho\sqrt{g_{\rho\rho}|_{\rho=0}} \,. 
		\end{equation}
		Thus, near $\rho=0$ one has $g_{\phi\phi}\left(\rho,z\right)\simeq g_{\rho\rho}\left(0,z\right)\rho^2+...$
	}:
	\begin{equation}
		g_{\phi\phi}\left(\rho,z\right)\simeq g_{\rho\rho}\left(0,z\right)\rho^2+...\,.
	\end{equation}
	Moreover, regularity of the Ricci scalar imposes that
	\begin{equation}
		\lim_{\rho\rightarrow0}\frac{g_{t\phi}}{g_{\phi\phi}}<\infty\,,
	\end{equation}
	that is, $g_{\phi\phi}$ cannot tend to zero faster than $g_{t\phi}$ in the axis limit~\cite{JaumeCarot_2000, Cunha:2020azh}.
	
	The remaining metric components must be finite at the axis. Thus, when approximating the axis the effective potentials take the form
	\begin{equation}
		H_{\pm}\sim\pm\sqrt{-\frac{g_{tt}\left(0,z\right)}{g_{\rho\rho}\left(0,z\right)}}\frac{1}{\rho}\,,
	\end{equation}
	from which it becomes clear that 
	\begin{equation}
		v^\pm_{\rho}\sim\mp\frac{1}{\rho^{2}}\,,v^\pm_{z}\sim\pm\frac{1}{\rho}\,.
	\end{equation}
	Hence
	\begin{align}
		\lim_{\rho\rightarrow0}\frac{v_{\rho}^{\pm}}{\left|{\bf v}^{\pm}\right|} & =\mp1\,,\\
		\lim_{\rho\rightarrow0}\frac{v_{z}^{\pm}}{\left|{\bf v}^{\pm}\right|} & =0\,.
	\end{align}
	Meaning that at the axis ${\bf v}^{\pm}$ will always be normal to it, and coincides with the vector field in the asymptotic limit. This means that ${\bf v}^{\pm}$ is constant along the segment ${\cal I}_1$, meaning that $w_{I_1}=0$. This, together with the result from the previous section implies that $w_\mathcal{I}=-w_\mathcal{H}$, \emph{i.e.} any contributions must come from the behaviour of the potentials near the horizon.
	
	\subsubsection{Horizon limit}
	
	Near the horizon we can define a set of Gaussian normal coordinates, $\left\{ X,Y \right\} $, such that $g_{XX}=1$ and $X|_{{\cal H}}=0$
	\footnote{Gaussian normal coordinates are a convenient set of coordinates naturally adapted to some hypersurface, $\mathcal{S}$, with normal vector $k$. To construct these coordinate system we consider at each point $p\in\mathcal{S}$ (parameterized by some coordinates $\{u^1,u^2,u^3\}$ on $\mathcal{S}$) the geodesic whose tangent vector is precisely $k$. Then, denoting the affine parameter along these geodesics as $X$, such that $X|_{\mathcal{S}}=0$, each point in the neighborhood of $\mathcal{S}$ has coordinates $\{u^1,u^2,u^3,X\}$. Eventually these coordinate system may become ill defined if the geodesics focus and cross, but they will always be valid in some neighborhood of $X=0$. For a deeper discussion see~\cite{Wald:1984rg} or~\cite{Carroll:2004st}.
	}. Near the horizon the potentials are then given by~\cite{Medved:2004tp,Cunha:2020azh}
	\begin{equation}
		H_{\pm}=\omega\pm\frac{\sqrt{D}}{{g_{\phi\phi}}}\simeq\omega\left(X=0,Y\right)\pm\frac{\kappa X}{\sqrt{g_{\phi\phi}^{{\cal H}}}}\,,
	\end{equation}
	where a possible $Y$ dependence of $\omega|_{X=0}$ was introduced. This means that 
	\begin{equation}
		\left.\partial_{X}H_{\pm}\right|_{X=0}=\pm\frac{\kappa}{\sqrt{g_{\phi\phi}^{{\cal H}}}}\,.
	\end{equation}
	Now we analyze $\partial_{Y}H_{\pm}|_{{\cal H}}$. Due to the toroidal topology of the event horizon it contains no fixed points of the $SO\left(2\right)$ group, hence $g_{\phi\phi}^\mathcal{H}>0$ (\emph{i.e.} it never vanishes), so 
	\begin{equation}
		\left.\partial_{Y}H_{\pm}\right|_\mathcal{H}=\frac{\partial\omega|_{X=0}}{\partial Y}\,,
	\end{equation}
	but regularity of the Ricci scalar on the horizon imposes that 
	\begin{equation}
		\omega_{H}:=\lim_{X\rightarrow0}\omega
	\end{equation}
	is constant and corresponds to the angular velocity of the horizon~\cite{Medved:2004tp}. So one obtains
	\begin{equation}
		\partial_{Y}H_{\pm}=0\,.
	\end{equation}
	Meaning that the vector field ${\bf v}$, in local coordinates on the horizon, points radially inwards or outwards (with respect to the new radial coordinate $x$), which means that its winding number when going around ${\cal H}^*$ is $w_\mathcal{H}=+1$. This manifests the fact that the horizon is a level surface of $H_\pm$. For topologically spherical BHs the horizon is also a level surface of $H_\pm$ with the exception of the poles, wherein the behaviour of $H_\pm$ is not regular.
	
	For a differential geometry proof of this statement notice that, since $\mathbf{v}$ is normal to $\mathcal{H}$ it will make a constant angle, $\pi/2$, with the tangent vector to $\mathcal{H}$, denoted $\mathbf{t}$. Therefore $\mathbf{v}$ and $\mathbf{t}$ will have the same winding number. And it is known that the winding number of the tangent vector of positively oriented closed curves is $+1$, see for example Sec. 1.7 of~\cite{Carmo}.
	
	Finally, one concludes that the total topological charge outside a toroidal event horizon is $w_\mathcal{I}=-1$. Therefore, there has to be at least one standard ($w=-1$) LR outside $\mathcal{H}$ per rotation sense ($i.e.$ for each $\pm$ sign of $H_\pm$). Additional ones must come in pairs with opposite topological charge, in order to preserve the total topological charge. It is worth noting that this result was reached without making use of any field equations, hence it is valid for any metric theory of gravity that satisfies the initial spacetime assumptions.
	
	\subsection{An illustrative example of a toroidal BH metric}
 
    Non topologically spherical BHs (in four dimensions that asymptotically approach Minkowski) are rather exotic and not common in the literature. In this sense, this subsection is devoted to a particular example of a line element containing a toroidal static BH. This metric is asymptotically Minkowskian (see App. (\ref{app:Eugen})), however its curvature decays slower than that of the Schwarzschild solution; thus, asymptotic flatness is not guaranteed in a rigorous sense - see e.g. chapter 11 of \cite{Wald:1984rg} or sec. 6.9 of \cite{Hawking:1973uf}. Nonetheless, it can be written in a coordinate system simultaneously adapted to both Killing vector fields, and the auxiliary vector field $\mathbf{v}$ has the correct asymptotic behavior. Therefore, since all other assumptions of the theorem are satisfied, this solution should respect it, \emph{i.e.} it should have a total topological charge $w=-1$ in the spacetime outside the event horizon.

	The considered metric was proposed in~\cite{KLEIHAUS2019134892},  and originally written in ring coordinates (see also~\cite{Emparan:2006mm}). Below we choose a slightly different coordinate system for the line element:

 	\begin{align}
		\mathrm{d}s^2&=-\frac{1+\lambda y}{\Lambda\left(\sigma,y\right)}\mathrm{d}t^2+\frac{R^2}{\left(\cos\sigma+y\right)^2}\Bigg[ \mathrm{d}\sigma^2\nonumber\\
		&+\frac{\left(1-\lambda \cos\sigma\right)^2}{\Lambda\left(\sigma,y\right)} \left(\frac{1}{1+\lambda y}\frac{\mathrm{d}y^2}{y^2-1} +\frac{y^2-1}{1-\lambda}\mathrm{d}\phi^2 \right)\Bigg]\,.\label{eq:Eugen}
	\end{align}
where $-\pi<\sigma\leq\pi$ and $y\in]-\infty,-1]$, with the coordinates $t$ and $\phi$ representing the usual time and azimuthal angular coordinates. See App. \ref{sec:TorCoord} for a discussion on the connection of this coordinate system to toroidal coordinates. Spatial infinity in this coordinate system corresponds to the point $(y=-1,\sigma=0)$, where the metric (\ref{eq:Eugen}) is singular, but this is merely a coordinate singularity which can be removed by an appropriate change of coordinates, see App. \ref{app:Eugen}.  The term $\Lambda\left(\sigma,y\right)$ is a strictly positive and smooth function that determines the far field behaviour of the spacetime. Following~\cite{KLEIHAUS2019134892} we can make the following explicit choice:
\begin{equation}
\Lambda(\sigma,y) = (1-\lambda)\left(1 + \frac{\sqrt{2}M}{R}\sqrt{-\cos\sigma -y}\right)\,,
\label{eq:LambdaChoice}
\end{equation}
where $M$ is the total (Komar) mass, which will be set to unity unless stated otherwise.

In addition, $R>0$ and $0<\lambda<1$ are free parameters related to the horizon ring size scale and location. The horizon in particular is located at $y=-1/\lambda$, which restricts the range of the $y$ coordinate to $\left]-1/\lambda,-1\right]$ inside the domain of outer-communication.

The horizon is a closed 2D orientable surface with vanishing Euler characteristic and $S^1\times S^1$ topology, \emph{i.e.} it is a torus~\footnote{There are surfaces with zero Euler characteristic which are no tori, for example the Klein bottle or the M{\"o}bius strip}. The Hawking temperature of the horizon is finite and non-zero for $0<\lambda<1$:
	\begin{equation}
		T_H =\frac{\sqrt{1-\lambda^2}}{4\pi R \lambda}\,.
	\end{equation}
	
	Using standard methods, \emph{e.g.}~\cite{Cunha:2022nyw}, it is possible to show that the metric~\eqref{eq:Eugen} has always a LR on the equatorial plane ($\sigma=0$) outside the horizon, located at:
	\begin{equation}
		y_{LR}=1-\sqrt{2+\frac{2}{\lambda}}\,.
	\end{equation}
	It is interesting to note that the location of the LR is independent of the choice for $\Lambda\left(\sigma,y\right)$. This result is consistent with toroidal BH spacetimes having at least one standard LR (provided they satisfy the previously stated assumptions).
	
	To further illustrate our result, we also computed the effective potentials $H_\pm$ for the metric (\ref{eq:Eugen}). The corresponding vector field $\mathbf{v}^+$ is plotted in Fig.~(\ref{fig:Eugen_flow}), for both the flat space limit $(\lambda=0\,,M=0)$ (top panel), and for a BH toroidal metric with $(\lambda=0.1\,,R=1\,,M=1)$ (bottom panel). These figures serve to illustrate how the boundary conditions at the event horizon are fundamental to the existence of a LR, namely how the vector field is normal to the horizon, leading to a saddle point of the effective potential $H_+$. A closer look near the critical point for the BH metric is provided in Fig. (\ref{fig:Eugen_zoom_flow}), where it is possible to better observe the circulation of the vector field around the critical point.

	\begin{center}
		\begin{figure}
			\begin{centering}
				\includegraphics[width=0.5\textwidth]{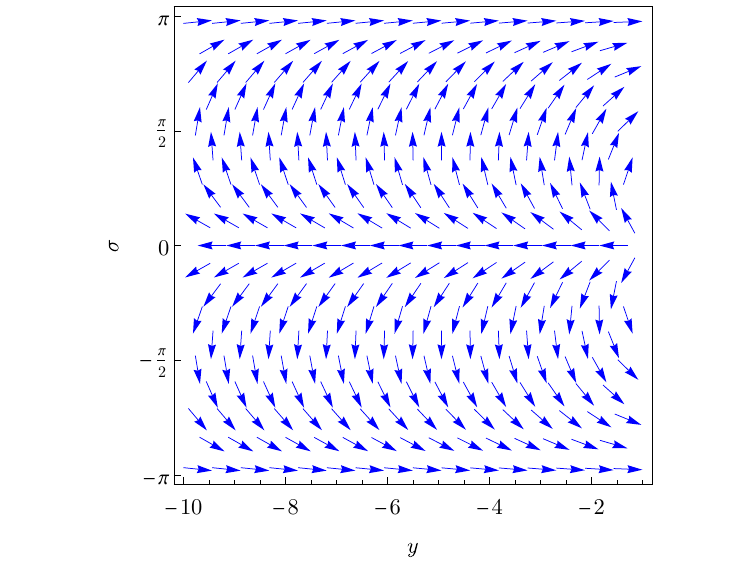}
				
				\includegraphics[width=0.5\textwidth]{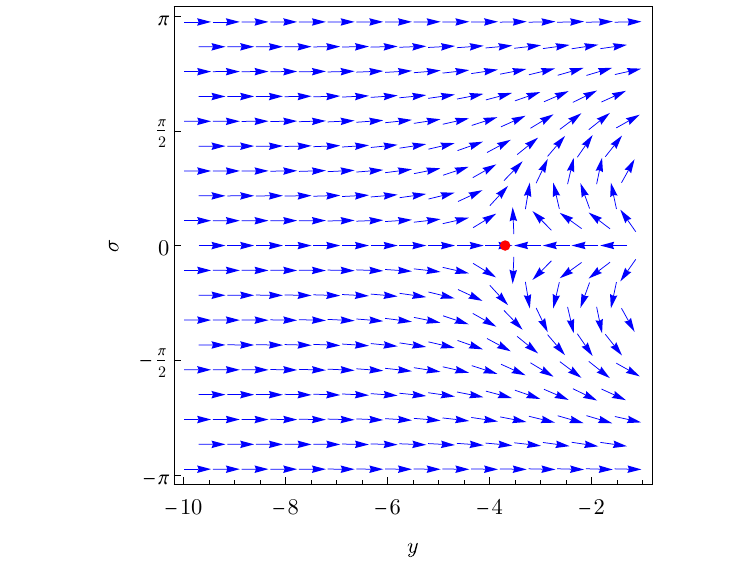}
				\par\end{centering}
			\caption{Vector field ${\bf v}$ in toroidal coordinates on flat spacetime and the spacetime of Eq. (\ref{eq:Eugen}). The LRs location is signaled with a red dot.}
			\label{fig:Eugen_flow}
		\end{figure}
		\par
	\end{center}

	\begin{center}
		\begin{figure}
			\begin{centering}
				\includegraphics[width=0.5\textwidth]{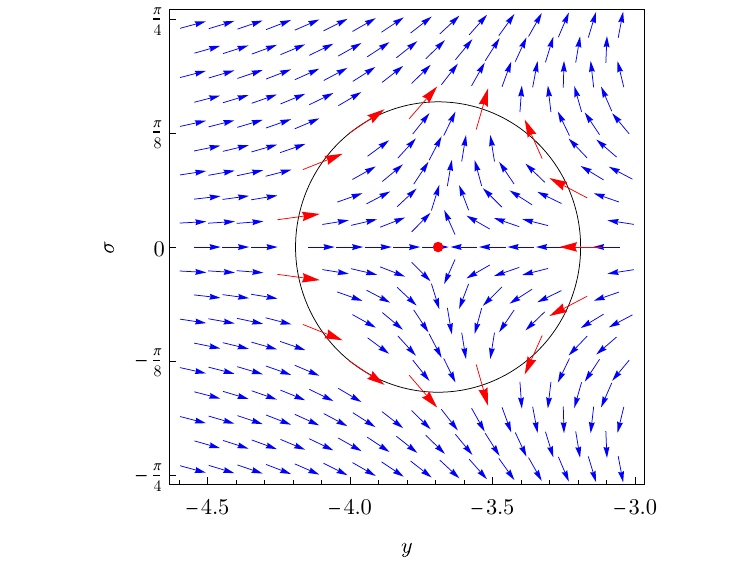}
				\par\end{centering}
			\caption{A closer look at the vector field in the vicinity of the LR. Here it is clearer that when traversing the black contour in the counterclockwise direction, the vector field (shown in red) will rotate in the clockwise direction, hence it will have a winding number of $-1$, as previously established.}
			\label{fig:Eugen_zoom_flow}
		\end{figure}
		\par
	\end{center}

	To make a connection with cylindrical-like coordinates, we performed a coordinate transformation between toroidal and cylindrical coordinates as if in flat space, see App. \ref{app:Eugen}, which suffices to have a well defined effective potential. We display $H_+$ in Fig. (\ref{fig:Eugen_Contour}), where it is clear that the horizon is a level set and the LR is a saddle point: The latter is clearly depicted by the LR level curve, shown as a thick black line.
	
	\begin{center}
		\begin{figure}
			\begin{centering}
				\includegraphics[width=0.4\textwidth]{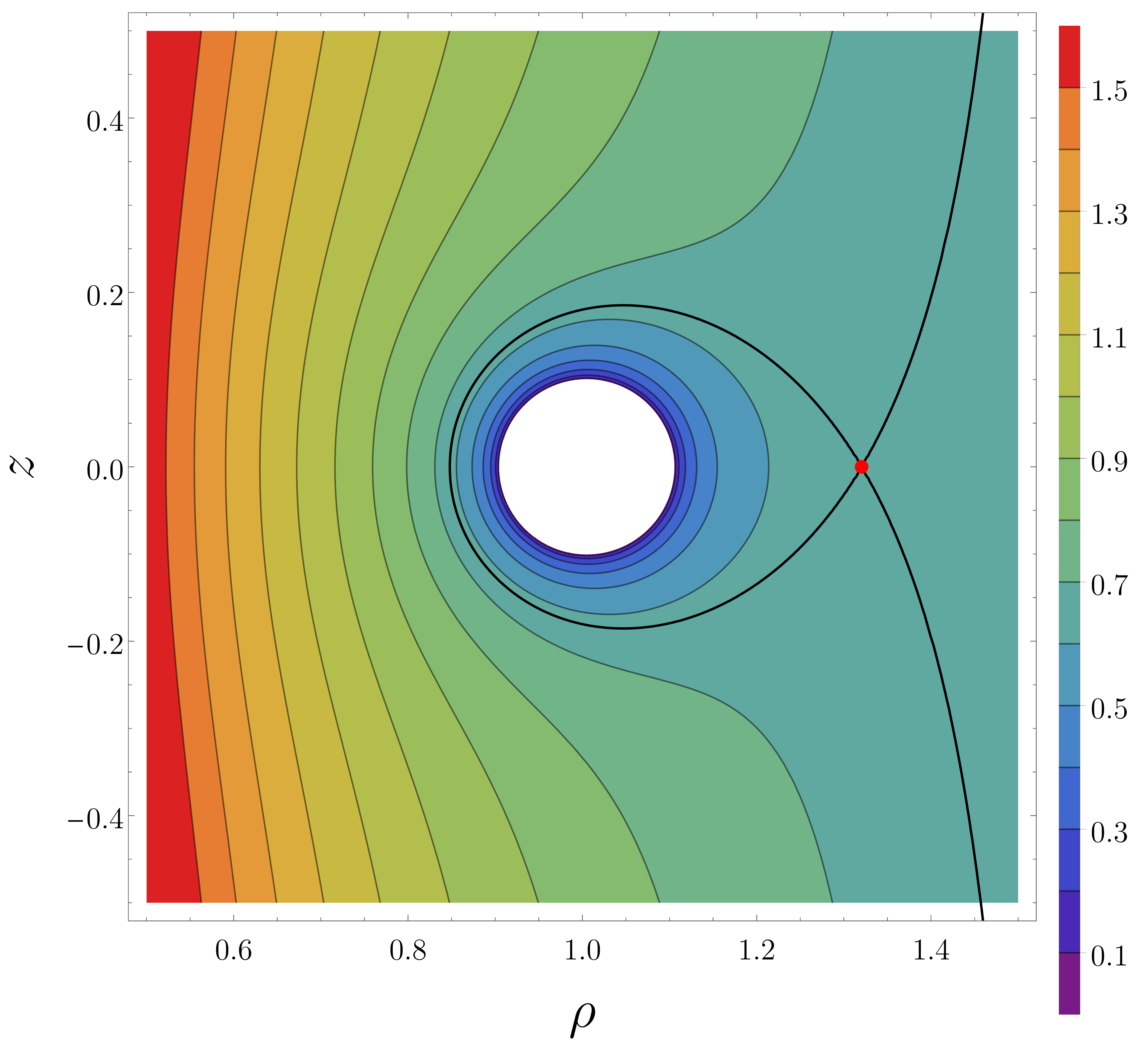}
				\caption{The contour plot of the effective potential $H_+$ for the metric (\ref{eq:Eugen}) in cylindrical-like coordinates, the saddle point, corresponding to the standard LR, is represented with a red dot. The level curve containing the critical point is represented with a thick black line.
					\label{fig:Eugen_Contour}}
			\end{centering}
		\end{figure}
		\par
	\end{center}
	
	Since four-dimensional BHs with toroidal topology are not commonly discussed in the literature, an additional analysis of the metric (\ref{eq:Eugen}) can be found in App. (\ref{app:Eugen}). For further illustration purposes, we display here some shadows of this toroidal BH in Fig. (\ref{fig:Eugen_Shadows}), where it is possible to appreciate how the topology of the horizon and the exotic features it produces has on these shadows. These images were obtained with a setup equivalent to the one described in~\cite{Cunha:2018acu,Cunha:2016bpi}. Some further details on computing these shadows are described in App. \ref{app:Imaging}.
	
	\begin{center}
		\begin{figure}
			\begin{centering}
				\includegraphics[width=0.23\textwidth]{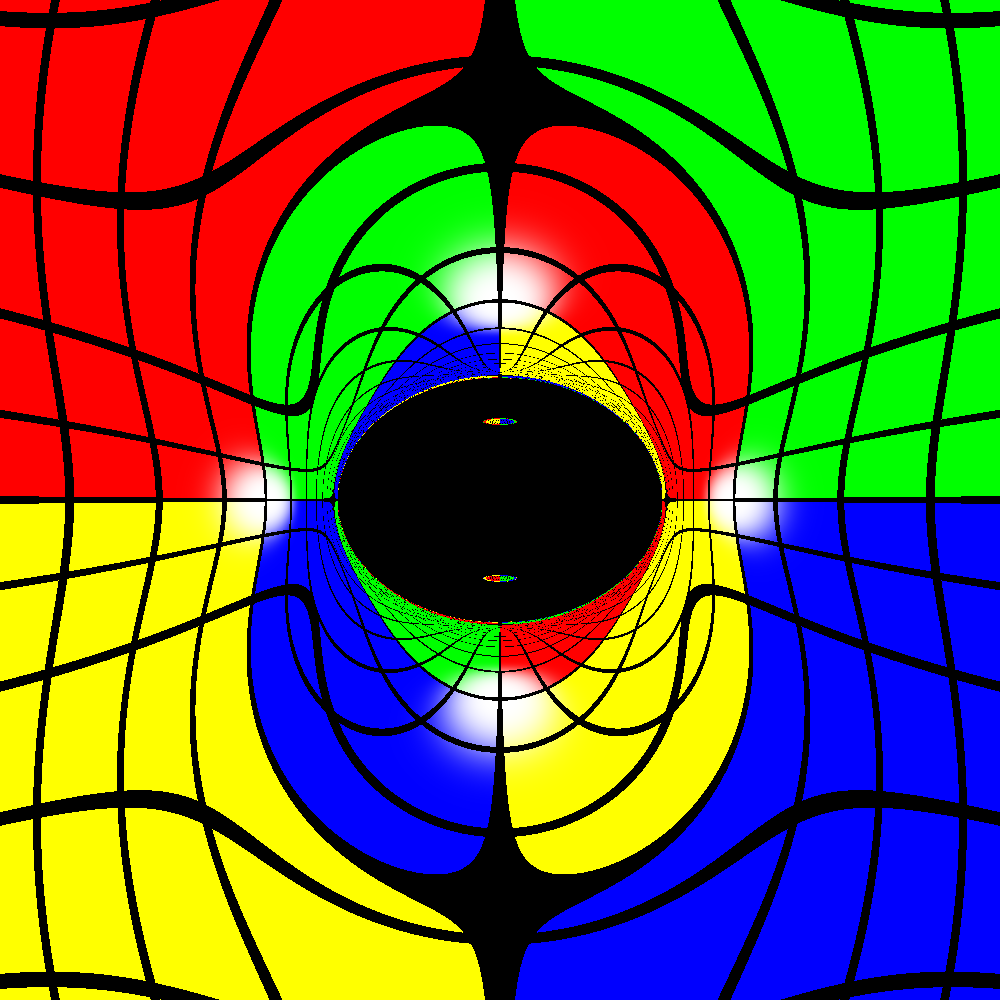}
				\includegraphics[width=0.23\textwidth]{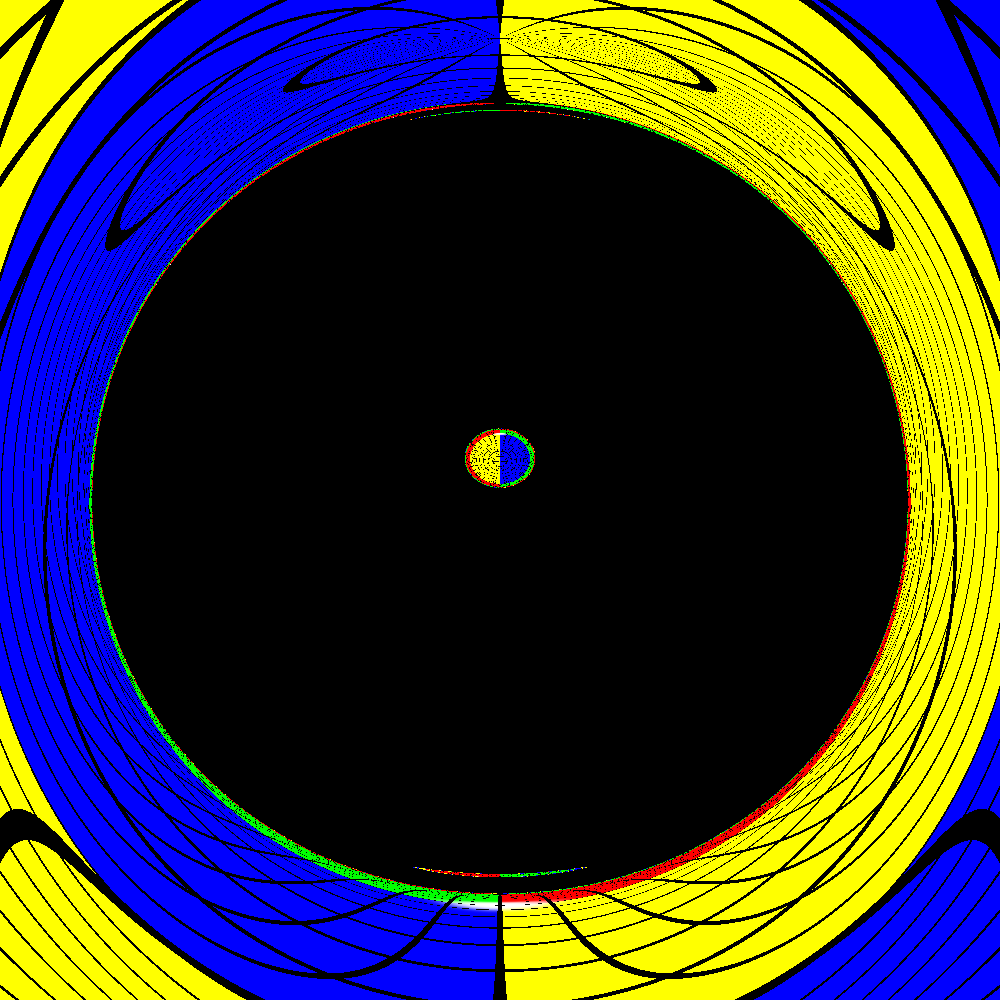}
				
				\vspace{3pt}
				\includegraphics[width=0.23\textwidth]{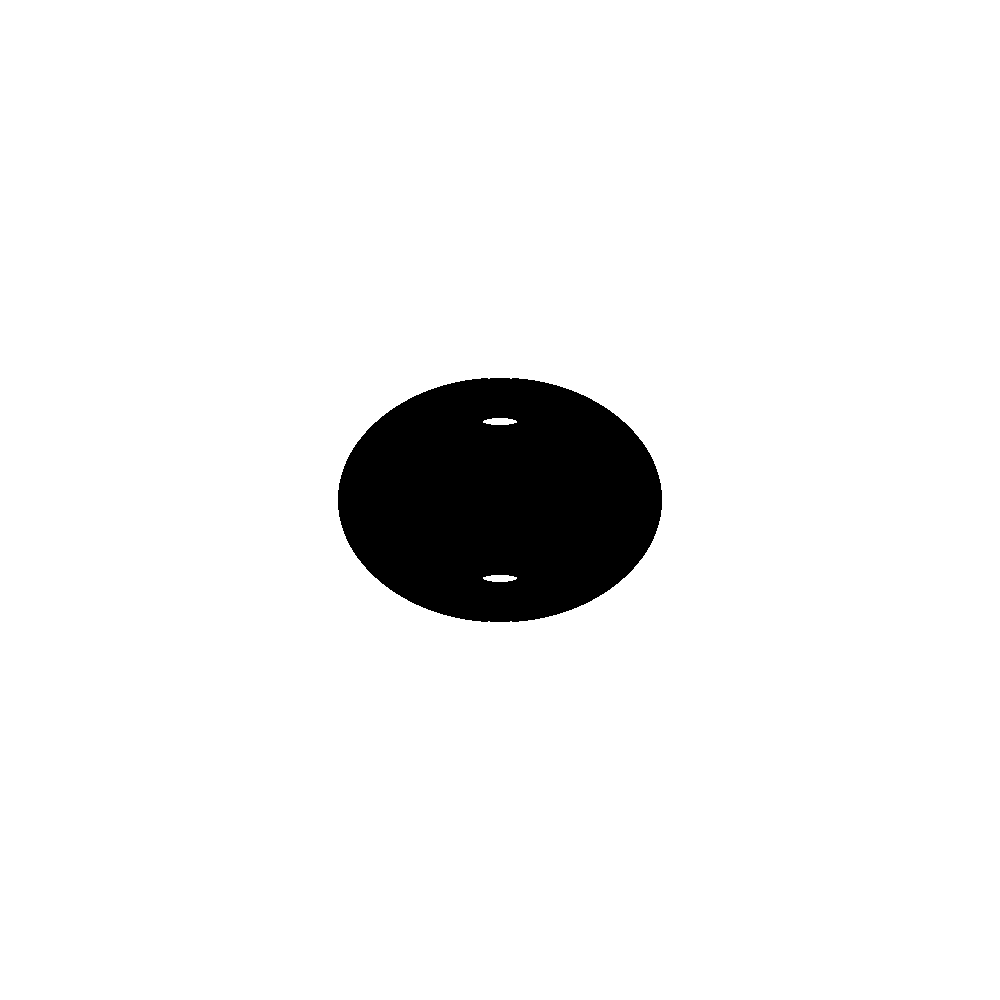}
				\includegraphics[width=0.23\textwidth]{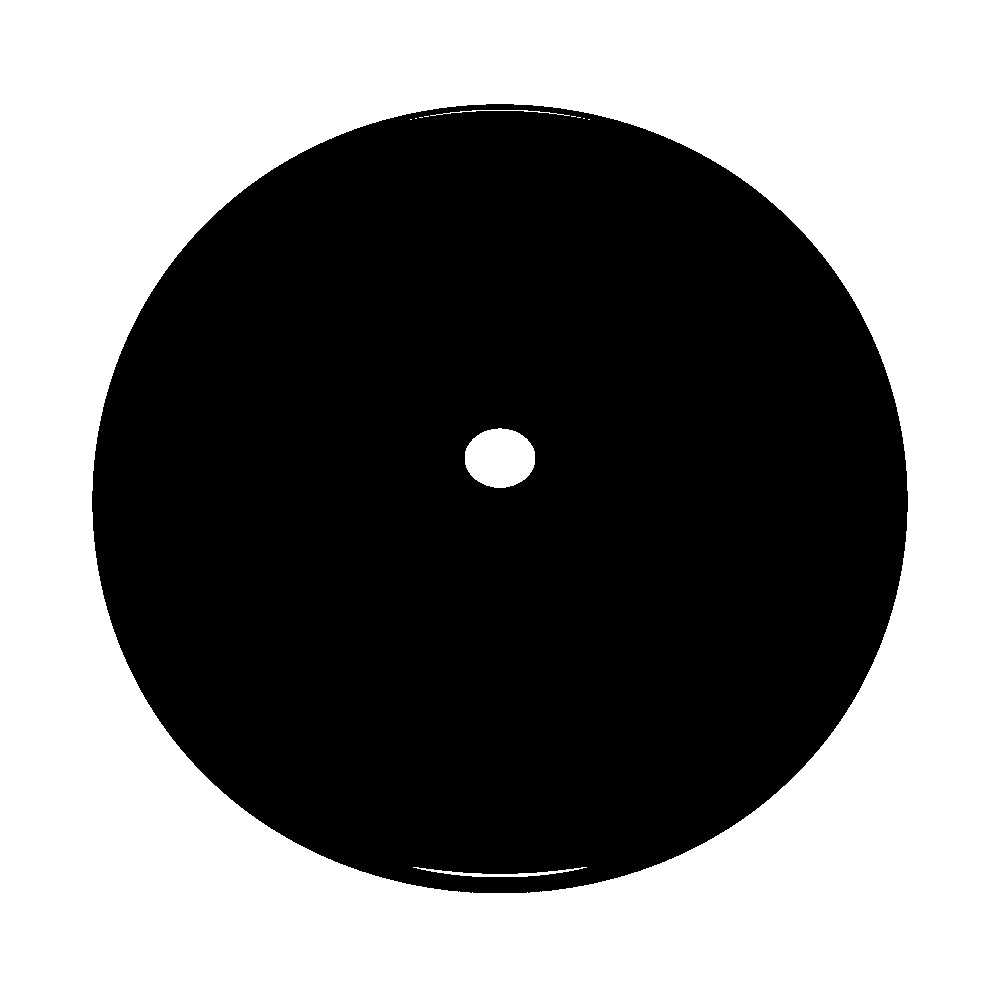}
				\par\end{centering}
			\caption{Shadows of the BHs described by Eq. (\ref{eq:Eugen}), with $\lambda=1/2\,,R=1\,,M=1$, for two different inclinations: on the left the observer lies on the equatorial plane, while on the right it has an inclination of $10$ degrees with respect to the equatorial plane. We refer the reader to App. \ref{app:Imaging} for more details on how the inclination is defined. On the top panel the images are obtained with a colored pattern on the celestial sphere, while on the bottom only the shadows is depicted.
				\label{fig:Eugen_Shadows}}
		\end{figure}
		\par
	\end{center}

	\section{Multiple toroidal BHs\label{sec:MultTor}}
	
	It is straightforward to generalise the previous construction to a spacetime possessing $n$ disconnected toroidal BHs, $\mathcal{H}_i\,,i=1,..,n$, all with a common axis, necessary in order to preserve axial symmetry. The orthogonal subspace of such configuration is represented in Fig. (\ref{fig:multiple_BH}), as well as the integration contour.
	
	At each event horizon one can construct a system of local Gaussian normal coordinates, meaning that the contribution of each toroidal event horizon to the total topological charge is $w_{\mathcal{H}_i}=-1$. Therefore, the total topological charge of the spacetime would be
	\begin{equation}
		w_\mathcal{E}=\sum_{i=1}^n w_{\mathcal{H}_i}=-n\,.
	\end{equation}
	This, in turn means that in such a spacetime there has to be at least one standard LR per BH (and rotation sense). Therefore, we can conclude that the existence of LRs are robust enough to the introduction of non-linear interactions between multiple toroidal BHs, regardless of their separation distance, so they obey a sort of \emph{superposition principle}.
 What if we consider multiple (topologically) spherical BHs? Such a case is considered in the following section.
	
	\begin{center}
		\begin{figure}
			\begin{centering}
				\includegraphics[width=0.45\textwidth]{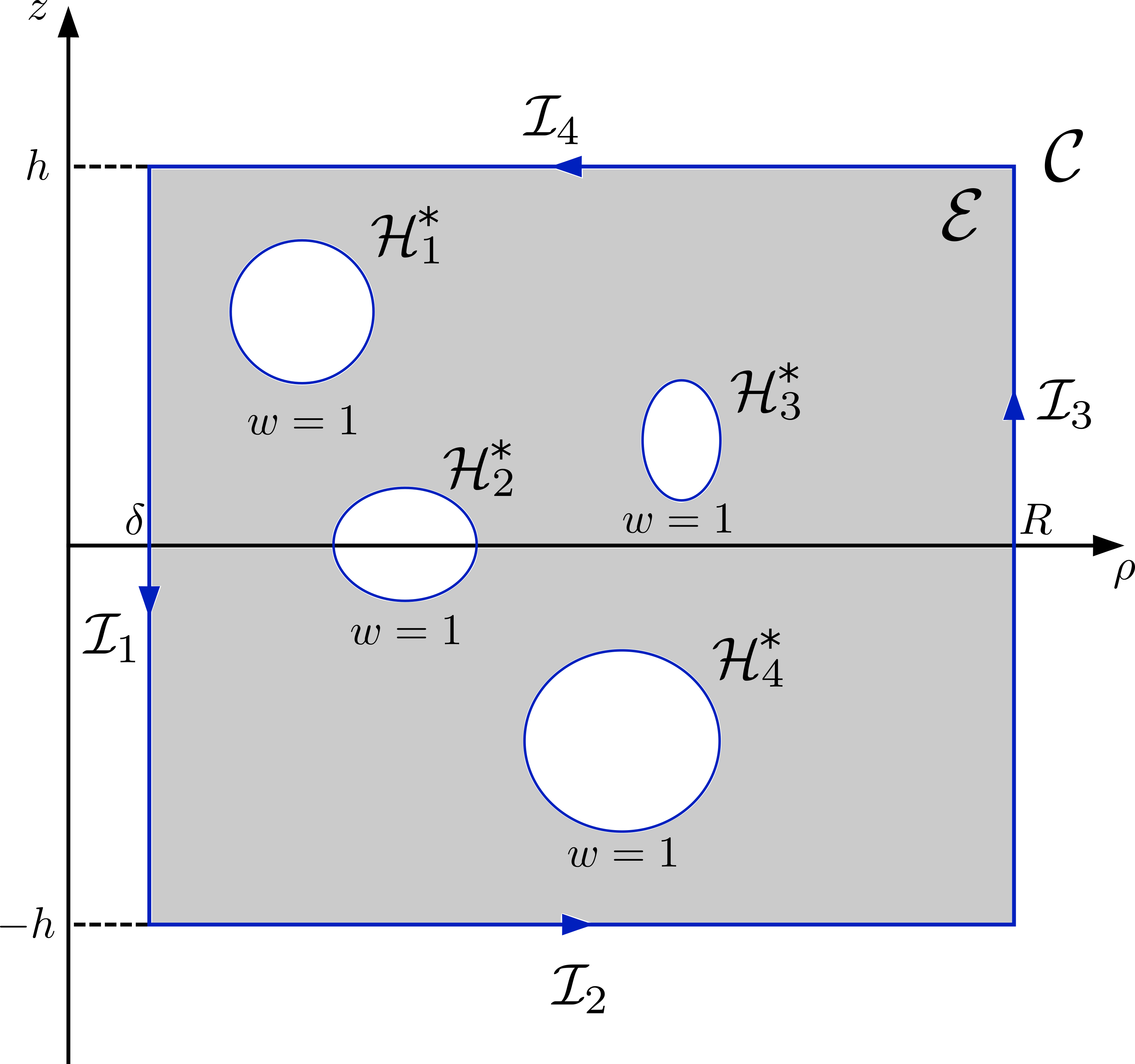}
				\par\end{centering}
			\caption{The orthogonal subspace of a spacetime harboring several disconnected toroidal event horizons. \label{fig:multiple_BH}}
		\end{figure}
		\par
	\end{center}
	
	\section{Multi collinear BH solutions\label{sec:MultiBHs}}
	
	Einstein's theory of relativity is exactly solvable for electrovacuum in $4$ dimensions if two commuting Killing vector fields exist~\cite{Weyl,Emparan:2001wk}. This can be used to construct analytical solutions, such as multi-BH spacetimes, known as Weyl solutions. The generation of these solutions is intimately connected to a set of boundary conditions defined on the $z$-axis, defined in terms of a rod structure~\cite{Harmark:2004rm}. This is due to a partial linearization of Einstein's equations, wherein some of the expressions reduce to Laplace-type equations, whose solution is determined by the rod structure. These Laplace-equations can regarded as the Poisson equation of Newtonian gravity where the sources lie along the $z$-axis.	Although this partial linearization is generically spoiled outside electrovacuum, the rod-structure setup can nevertheless be applied in the presence of matter content, as exemplified by the numerical construction of two BH solutions balanced by a scalar field in the literature, for both static~\cite{Herdeiro:2023mptSch} and rotating~\cite{Herdeiro:2023roz} configurations. 
	
	These constructions motivates the use of such rod structures on the $z$-axis for the metric \eqref{eq:LineEl}, in order to describe a spacetime with multiple collinear (topologically spherical) BHs in equilibrium, provided that the spacetime is still axial-symmetric and stationary. 
	
	To introduce the rod structure one should work in Bach-Weyl coordinates, $\left\{ t,\phi,\rho,z\right\} $, which are asymptotically cylindrical. The radial coordinate $\rho$ is defined as the square root of the modulus of the determinant of the Killing part of the metric, hence all points left invariant by the action of the Killing vector fields have $\rho=0$, \emph{i.e.} they lie along the $z$-axis. Since we are interested in multi BH solutions we consider fixed points of the spacelike vector,
	$\eta=\partial_{\phi}$, and of the null event horizon generator $\chi$. A necessary condition for the regularity of the solution is that that each
	point of the $z$-axis is a fixed point of precisely one of the Killing vector fields, except in isolated points, the \textit{common points}. Assuming that the spacetime contains $N$ aligned (topologically) spherical BHs, we can denote these common points as $\{a_{1},...,a_{2N}\}$. It is then possible to divide the $z$-axis into $2N+1$ intervals $\left[a_{i},a_{i+1}\right],i\in\{0,...,2N\}$, where $a_{0}:=-\infty$ and $a_{2N+1}:=+\infty$ are not common points. These line segments are known as the rods of the solution~\cite{Harmark:2004rm}.
	
	Each rod is either spacelike and it is part of the rotation axis, or it is timelike and is one of the BHs. For multi aligned BHs the semi-infinite rods $\left[a_{0},a_{1}\right]$ and $\left[a_{2N},a_{2N+1}\right]$ are spacelike~\footnote{Semi-infinite timelike rods describe an accelerating BH} and correspond to the rotation axis.
	
	To determine the number of LRs in such a spacetime we have to compute the winding number of the vector field along a contour
	which, in the appropriate limit, encompasses all the spacetime outside the event horizons. This contour is similar to the one had considered for the case of toroidal BHs, except now all BHs intersect the axis. Since we still assume asymptotic flatness the behaviour of the vector field in the asymptotic limit will be the same. Furthermore, it is assumed that there are no event horizons inside the contour, such that the total winding number is given by $w_\mathcal{E}=w_{I_1}+w_{I_2}+w_{I_3}+w_{I_4}=w_{I_1}$, since $w_{I_{2,3,4}}=0$ on account of asymptotic flatness.
	
	Therefore, we have to study the behaviour of the effective potentials $H_{\pm}$ near $\rho=0$. However, the behaviour along the $z$-axis will differ depending on whether we are at an event horizon or at the rotation axis, and these two cases must be studied carefully. If the metric is expressed in coordinates adapted to the Killing vector fields which vanish on the axis, namely $\eta=\partial_\phi$ and $\chi=\partial_u$, then it takes the following form near a spacelike rod (axis)~\cite{Harmark:2004rm}
	\begin{equation}
		{\rm d}s^{2}=-A\left(z\right){\rm d}u^{2}+\frac{1}{A\left(z\right)}\left[\rho^{2}{\rm d}\phi^{2}+c_{1}^{2}\left({\rm d}\rho^{2}+{\rm d}z^{2}\right)\right]\,, \label{eq:SpaRod}
	\end{equation}
	and near a timelike rod (BH)
	\begin{equation}
		{\rm d}s^{2}=B\left(z\right){\rm d}\phi^{2}+\frac{1}{B\left(z\right)}\left[-\rho^{2}{\rm d}u^{2}+c_{2}^{2}\left({\rm d}\rho^{2}+{\rm d}z^{2}\right)\right]\,.\label{eq:TimRod}
	\end{equation}
	Here $A$ and $B$ are strictly positive functions of $z$. The absence of a cross term means that these are locally co-rotating coordinates.
  In order to avoid conical singularities at the axis, the period of the angular coordinate $\phi$ needs to be $2\pi c_1$, which implies $c_1=1$ given the usual $2\pi$ azimuthal angle identification. Curiously, we can relate this analysis close to the rotation axis with the one close to a BH, since the form of the two line elements~\eqref{eq:SpaRod}-\eqref{eq:TimRod} is actually very similar. This point can be made clearer by Wick rotating the timelike coordinate $u$ into a spacelike coordinate $\tilde{u}=\mathrm{i}u$. By then  applying a similar reasoning to $\tilde{u}$ as the one performed in $\phi$, one concludes that the coordinate $\tilde{u}$ needs to have a period of $2\pi c_2$. This periodicity is actually related to the Hawking temperature of the horizon via $T_H=\left(2\pi c_{2}\right)^{-1}$. Since we will only consider non-extremal BHs, then  $0<c_{2}<\infty$, which implies that the local line element expansion (\ref{eq:TimRod}) is not well defined for extremal BHs.
	
	Since these coordinate systems, at each rod, are locally co-rotating near the axis the effective potentials reduce to
	\begin{equation}
		H_\pm\simeq\pm\sqrt{-\frac{g_{uu}}{g_{\phi\phi}}}\,,
	\end{equation}
	where the behaviour of the metric functions will depend on the nature of the rod. Each case will be considered separately below, as well as the behaviour at the common points. For clarity sake, we shall focus below on the potential $H_+$, but a virtually identical analysis applies to $H_-$.
	
	\subsection{Spacelike rods (axis)}
	
	Near a spacelike rod (axis) the potential goes like
	\begin{equation}
		H_{+}\sim\frac{A\left(z\right)}{\rho}\,.
	\end{equation}
	Thus, since $A(z)\neq0$ (otherwise we would be at a common point):
	\begin{equation}
		{\bf v}=(v_\rho,v_z)\sim\left(-\frac{1}{\rho^{2}},\frac{1}{\rho}\right)\,,
	\end{equation}
	which implies that the normalized vector field at a spacelike rod is
	\begin{equation}
		\left.\frac{{\bf v}}{v}\right|_{\rho=0}=\left(-1,0\right)\,.
	\end{equation}
	Thus, at a spacelike rod the vector fields points towards the rotation axis (which is consistent with the previous conclusion in the toroidal BH section). It should be noted that the vector coincides with the asymptotic limit behaviour, hence the only non-vanishing contribution to the topological charge will come from the BHs.
	
	\subsection{Timelike rods (BH)}
	
	Near a timelike rod (BH)
	\begin{equation}
		H_{+}\simeq\frac{\rho}{B\left(z\right)}\,.\label{eq:Htimerod}
	\end{equation}
	Once again we do not consider the common point, therefore $B(z)\neq0$, meaning that at the event horizons
	\begin{equation}
		\left.\frac{{\bf v}}{v}\right|_{\rho=0}=\left(1,0\right)\,.
	\end{equation}
	Thus, at an event horizon the vector fields points outwards. This correspond to a difference of $\pi$ when compared with the field at the rotation axis. But to compute the winding number it is necessary to know how this phase of $\pi$ is obtained when crossing the common points. To determine how the vector field rotates at the common points it is necessary to have a more detailed analysis and to proceed with care, as discussed below.

	\subsection{Common points (BH poles)}
		
	In this section we will study the behaviour of $\mathbf{v}$ when approaching the common points either from the timelike (BH) or spacelike (axis) rods. In doing so we consider $\rho\ll1$ but $\rho\neq0$, such that the expansions (\ref{eq:TimRod}) and (\ref{eq:SpaRod}) can still be considered valid, and the derivatives $A^\prime$ and $B^\prime$ are well defined (these derivatives are not well defined at the common points).
	
	We start by considering a section of a spacelike rod (axis) such that the vector field reads:
	\begin{equation}
	{\bf v}=\left( 
	-\frac{A\left(z\right)^{3/2}}{\rho^2},\frac{\sqrt{A\left(z\right)}}{\rho}A^\prime\left(z\right)
	\right)\,.
	\end{equation}
	Notice how $v_\rho<0$, \emph{i.e.} the vector field is always pointing towards the axis. Regularity of the Ricci scalar also implies that near a common point one has: $A\left(z\right)\simeq \pm (z-z_*)^n\,,n\geq2$ (here $z_*$ denotes a common point), with the sign determined by both the parity of $n\in\mathbb{N}$ and the location of the rod with respect to $z_*$ ( $i.e.$ whether it is above or below the latter). However, the limiting behaviour to the common point is unaffected by the specific power $n$ considered. For the sake of simplicity, we shall focus on the simplest case $n=2$ in our computations, while keeping in mind that the results are valid for any $n>2$. Using the regularity restriction above, the normalized vector fields reads:
	\begin{equation}
		\frac{\mathbf{v}}{v}\simeq\left(\frac{\left|z-z_*\right|}{\sqrt{4 \rho ^2+(z-z_*)^2}}, \frac{2 \rho  \left| z-z_*\right| }{(z-z_*) \sqrt{4 \rho ^2+(z-z_*)^2}}\right)\,.
	\end{equation}
	Therefore, it is clear that in the limit of approaching the common point $z_{*}$ from the axis side:
	\begin{equation}
		\frac{\mathbf{v}}{v}=\left(0,\pm1\right)\,,\quad z\rightarrow z_{*}^{\pm}\,, \label{eq:LimSpa}
	\end{equation}

 where $z_*^+$ ($z_*^-$) denotes whether $z_{*}$ is approached from positive (negative) values of the coordinate $z$.
	Equation~\eqref{eq:LimSpa} is then stating that if $z_*$ is being approached from above ($z>z_*$) then $\mathbf{v}$ points upwards. However, if $z_*$ if approached from below ($z<z_*$), then $\mathbf{v}$ points downwards. Since $\mathbf{v}$ points towards the axis when very near the rod, then these limits correspond to a rotation of $-\pi/2$ and $+\pi/2$, respectively, as the common point is reached from the axis side.\\
	
	Consider now that we approach $z_*$ from the side of a timelike rod (BH). Then the vector field reads:
	\begin{equation}
		\frac{\mathbf{v}}{v}\simeq\left(\frac{1}{\sqrt{c_2 B(z)}},-\frac{\rho  B'(z)}{\sqrt{c_2}
		B(z)^{3/2}}\right)\,.
	\end{equation}
	Notice now how $v_\rho>0$, \emph{i.e.} the vector fields points always away from the axis. The function $B$ is subject to the same regularity conditions as $A$, so we assume $B\simeq \left(z-z_*\right)^2$ and obtain:
	\begin{equation}
		\frac{\mathbf{v}}{v}\simeq\left(\frac{\left|
			z-z_*\right|}{\sqrt{4 \rho ^2+\left(z-z_*\right){}^2}  }, -\frac{2 \rho  \left(z-z_*\right)}{\left| z-z_*\right|  \sqrt{4 \rho
				^2+\left(z-z_*\right){}^2}} \right)\,.
	\end{equation}
	Therefore, 
	\begin{equation}
		\frac{\mathbf{v}}{v}=\left(0,\mp1\right)\,,z\rightarrow z_{*}^{\pm}\,.\label{eq:LimTim}
	\end{equation}
	
This conclusion is actually consistent with the results of (\ref{eq:LimSpa}): the limiting approach to a common point {\it from below} in the neighbourhood of a spacelike rod (axis) must be compared with the limit to that same common point but {\it from above} near a timelike rod (BH), and vice versa. This comparison is important, since the field $\mathbf{v}$ should have a well defined limit at each common point, regardless of the rod it is being approach from.\\

 Let us examine the path along the contour segment that runs down through the $z$-axis. We begin at a spacelike rod (axis) where the vector points towards the axis. As we progress, we eventually arrive at a common point, which is approached from above. As this common point is approached, according to equation (\ref{eq:LimSpa}), the vector points upwards and has rotated by $-\pi/2$. Moving beyond this common point, we find ourselves near a timelike rod (BH) where the vector field must point away from the axis, with $v_\rho>0$. Consequently, it will undergo an additional rotation of $-\pi/2$. This implies that the total circulation of the vector when crossing this initial common point is $-\pi$. It is important to note that, based on our assumption, all timelike rods (BHs) are finite as they extend down the axis. Thus, we will encounter a new common point further down. At this subsequent common point, the vector transitions from pointing away from the axis to pointing downwards, as dictated by (\ref{eq:LimTim}), corresponding to a rotation of $-\pi/2$, since $v_\rho>0$. Upon crossing this common point, we return to a spacelike rod (axis), where the vector field changes from pointing downwards to pointing towards the axis with $v_\rho<0$. Consequently, the vector field gains an additional winding of $-\pi/2$, totalling $-\pi$ after crossing this common point, similar to the previous transition from a spacelike to timelike rod. This analysis allows us to conclude that the winding number associated with a single timelike rod (representing a single topologically spherical BH), delimited by two common points, is $w=-1$, since the vector $\mathbf{v}$ undergoes a rotation of $-2\pi$.

	\subsection{Total topological charge}
	
	Since the first and last rods are spacelike and we have $N$ horizons, then one has $2N$ common points, and the vector field rotates $2N\pi$ clockwise. Since the contour is being traversed in the counterclockwise direction this gives a contribution to the winding number of 
	\begin{equation}
		w_{\mathcal{I}_1}=\frac{1}{2\pi}\oint_{{\cal I}_1}{\rm d}\Omega=-N\,.
	\end{equation}
	As discussed previously asymptotic flatness implies that this is the only contribution to the winding number, then the total winding number is 
	\begin{equation}
		w=\frac{1}{2\pi}\oint_{{\cal C}}{\rm d}\Omega=-N\,.
	\end{equation}
	
	Since the total topological charge counts the number of LRs we conclude that there is at least one standard LR per BH (and rotation sense). So, similarly to the case of multiple toroidal BHs, LRs around spherical BHs are robust against the non-linear interaction between multiple BHs and coexist regardless of the separation between the horizons.
	
	As an illustration of this result we plot the vector field $\mathbf{v}$ for an analytical solution containing two event horizons in equilibrium obtained in~\cite{Chen:2012dr} which satisfies all our assumptions. This is a Kaluza-Klein solution corresponding to the dimensional reduction of a five-dimensional vacuum solution of Einstein's equations. The two BHs in the solution are either purely electric or purely magnetic and are balanced by the dilaton field. This solution is composed of 5 rods: $3$ spacelike and $2$ timelike ($i.e.$ the BHs), which are determined by $4$ fixed common points. Additionally, one needs to specify two extra points along the axis to fully specify a particular solution in this model. Thus, in our notation, the solution is specified by a set $\{z_1,a_1,a_2,a_3,a_4,z_2\}$.
	
	\begin{center}
		\begin{figure}
			\begin{centering}
				\includegraphics[width=0.45\textwidth]{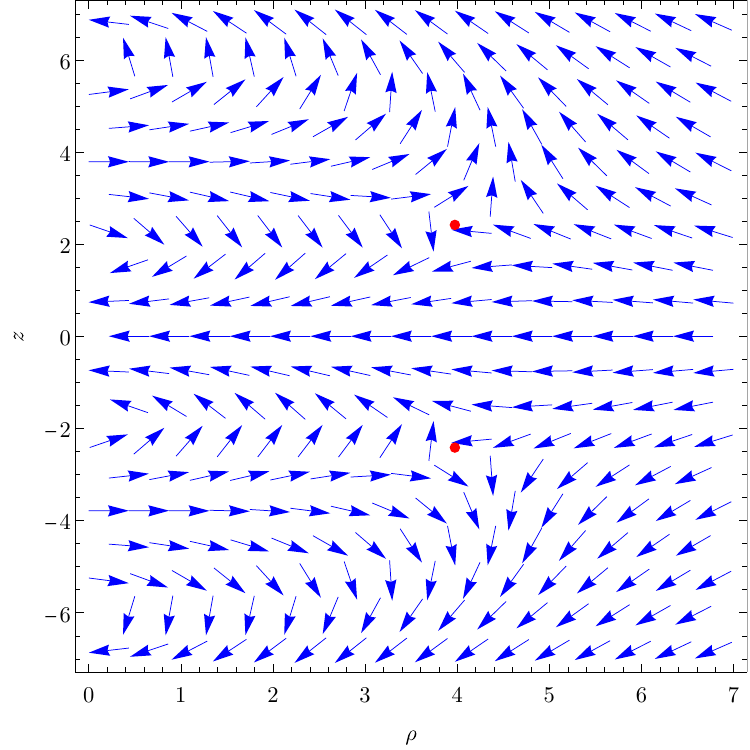}
				
				\includegraphics[width=0.45\textwidth]{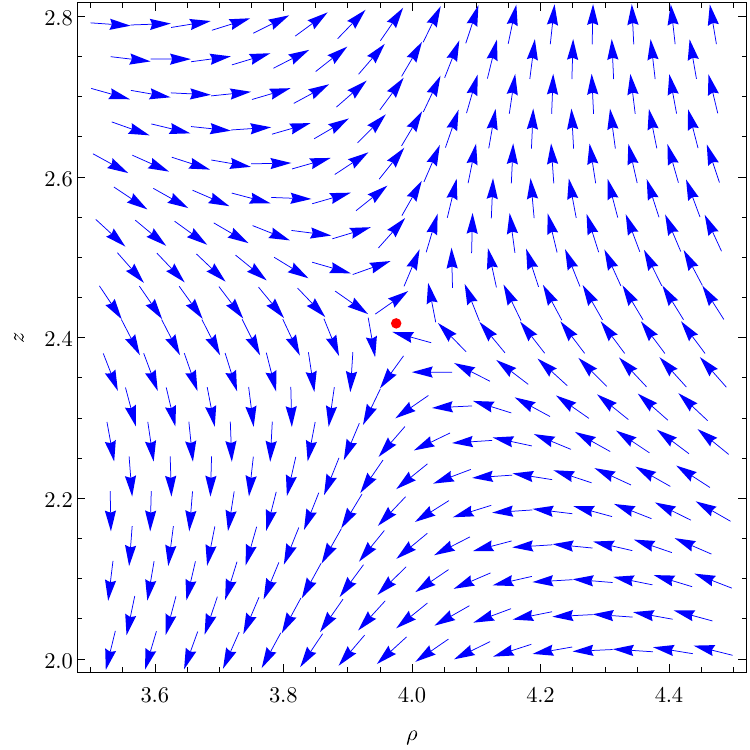}
				\par\end{centering}
			\caption{Vector field ${\bf v}$ for the double BH spacetime of~\cite{Chen:2012dr} with parameters $z_1=-7$, $a_1=-6$, $a_2=-2$, $a_3=2$, $a_4=6$, $z_2=7$. The critical points, corresponding to the LRs are represented as the red points. The bottom panel corresponds to a region close to the "upper" LR.
				\label{fig:Teo_flow}.}
		\end{figure}
		\par
	\end{center}
	
	In Fig. (\ref{fig:Teo_flow}) it is possible to verify that the vector field has two critical points, as expected from the result derived earlier. Moreover, it is possible to appreciate the change on its behaviour along the $z$-axis depending on the nature of the rod, pointing outwards for event horizons and inwards for the rotation axis. In the bottom panel it is clear that the circulation of the vector field around the critical points yields $\omega=-1$, corresponding to a standard LR (the same occurs for the other LR).
	
	\section{Combining topologically spherical and toroidal BHs\label{sec:Everything}}
	
	Since the rod structure can coexist with toroidal BHs, which do not intersect the symmetry axis, it means that it is possible to combine the two previous results with multiple BHs into a single theorem: consider a $3+1$-dimensional spacetime that is asymptotically flat, stationary and circular that is harboring $n$ toroidal BHs and $N$ topologically spherical BHs, all sharing a common axis of symmetry. Then, as implied by the previous results, there will be at least one standard LR per BH, $i.e$ at least $n +N $ LRs. Any additional LRs, if they exist, must come in pairs with opposite charges. This result is curiously stating that LRs satisfy a {\it de facto} superposition principle, although the interaction between the BHs is fully non-linear.

	\section{Discussion and final remarks}
	
	In this work we have generalized the theorem put forward in~\cite{Cunha:2020azh} by dropping some of its assumptions, namely the number of BHs and their topology.
 
 Since we preserved the assumption of axial symmetry, the only other possible option for the topology of the horizon (beyond spherical) is toroidal topology. We first considered spacetimes with a single toroidal BH, and reached the conclusion that they satisfy a result similar to their spherical counterparts, \emph{i.e.} each BH must possess at least one standard LR per rotation sense, and if additional ones exist they must come up in pairs with opposite winding numbers. 

Is this result dependent on the number of BHs? It is intuitive that when two BHs are very far apart they interact weakly, and so it is not unexpected that each BH would have its own LR in that situation. However, when the BHs are much closer together, and non-linear interactions become relevant, it is non-trivial whether individual LRs would survive the interaction. However by further generalizing the result with multiple toroidal BHs it is possible to conclude that they actually do survive. So LRs obey a curious sort of superposition principle that holds in a non-linear regime.
	
	Motivated by this result, we then extended the original setup of~\cite{Cunha:2020azh} into a collection of collinear (and topologically spherical) BHs, all aligned along the symmetry axis. Once again it was possible to assert that each BH contributes with a standard LR, at least, to the spacetime.
	
	Finally, by combining both results, one can conclude that the superposition still holds regardless of the topology and number of the horizons, as long as the spacetime symmetries are maintained. Meaning that the each BH is intrinsically connected to a LR.
	
	As a byproduct of our analysis we also studied the line element proposed in~\cite{KLEIHAUS2019134892}, describing a spacetime with a toroidal BH satisfying the assumptions of the theorem. The study focused mainly on the nature of the horizon and on its shadow. The shadows produced here are to the authors knowledge the first ones produced for a fully 4-dimensional toroidal BH, despite lensing images of 5-dimensional toroidal BHs having already been reported~\cite{Hertog:2019hfb}.

 \bigskip
	
	\section{Acknowledgments}
	
	The authors want to express their gratitude to Eugen Radu for insightful discussions and comments during the writing of this paper. This work is supported by the Center for Research and Development in Mathematics and Applications (CIDMA) through the Portuguese Foundation for Science and Technology (FCT - Fundação para a Ciência e a Tecnologia), references https://doi.org/10.54499/UIDB/04106/2020 and https://doi.org/10.54499/UIDP/04106/2020. The authors acknowledge support from the projects http://doi.org/10.54499/PTDC/FIS-AST/3041/2020, as well as http://doi.org/10.54499/CERN/FIS-PAR/0024/2021 and https://doi.org/10.54499/2022.04560.PTDC. This work has further been supported by the European Union’s Horizon 2020 research and innovation (RISE) programme H2020- MSCA-RISE-2017 Grant No. FunFiCO-777740 and by the European Horizon Europe staff exchange (SE) programme HORIZON-MSCA-2021-SE-01 Grant No. NewFunFiCO101086251. J. N. is supported by the FCT grant 2021.06539.BD. PC is supported by the Individual CEEC program 2020 funded by the FCT, with DOI reference 10.54499/2020.01411.CEECIND/CP1589/CT0035. Computations have been performed
at the Argus and Blafis cluster at the U. Aveiro.

	\bibliographystyle{unsrt}
	\addcontentsline{toc}{section}{\refname}\bibliography{refs}

\begin{thebibliography}{10}

\bibitem{Cunha:2020azh}
Pedro V.~P. Cunha and Carlos A.~R. Herdeiro.
\newblock {Stationary black holes and light rings}.
\newblock {\em Phys. Rev. Lett.}, 124(18):181101, 2020.

\bibitem{Cardoso:2016rao}
Vitor Cardoso, Edgardo Franzin, and Paolo Pani.
\newblock {Is the gravitational-wave ringdown a probe of the event horizon?}
\newblock {\em Phys. Rev. Lett.}, 116(17):171101, 2016.
\newblock [Erratum: Phys.Rev.Lett. 117, 089902(E) (2016)].

\bibitem{Cunha:2018acu}
Pedro V.~P. Cunha and Carlos A.~R. Herdeiro.
\newblock {Shadows and strong gravitational lensing: a brief review}.
\newblock {\em Gen. Rel. Grav.}, 50(4):42, 2018.

\bibitem{hod2013upper}
Shahar Hod.
\newblock Upper bound on the radii of black-hole photonspheres.
\newblock {\em Physics Letters B}, 727(1-3):345--348, 2013.

\bibitem{paganini2018role}
Claudio~Francesco Paganini.
\newblock {\em The role of trapping in black hole spacetimes}.
\newblock PhD thesis, Universit{\"a}t Potsdam, 2018.

\bibitem{Junior:2021svb}
Haroldo C. D.~Lima Junior, Jian-Zhi Yang, Lu\'\i{}s C.~B. Crispino, Pedro V.~P.
  Cunha, and Carlos A.~R. Herdeiro.
\newblock {Einstein-Maxwell-dilaton neutral black holes in strong magnetic
  fields: Topological charge, shadows, and lensing}.
\newblock {\em Phys. Rev. D}, 105(6):064070, 2022.

\bibitem{Hawking:1971vc}
S.~W. Hawking.
\newblock {Black holes in general relativity}.
\newblock {\em Commun. Math. Phys.}, 25:152--166, 1972.

\bibitem{hawking_ellis_2023}
Stephen~W. Hawking and George F.~R. Ellis.
\newblock {\em The Large Scale Structure of Space-Time: 50th Anniversary
  Edition}.
\newblock Cambridge Monographs on Mathematical Physics. Cambridge University
  Press, 2023.

\bibitem{Gannon1976}
Dennis Gannon.
\newblock {On the Topology of Spacelike Hypersurfaces, Singularities, and Black
  Holes}.
\newblock {\em Gen.Rel.Grav.}, 7(2):219, 1976.

\bibitem{Friedman:1993ty}
John~L. Friedman, Kristin Schleich, and Donald~M. Witt.
\newblock {Topological censorship}.
\newblock {\em Phys. Rev. Lett.}, 71:1486--1489, 1993.
\newblock [Erratum: Phys.Rev.Lett. 75, 1872(E) (1995)].

\bibitem{Krasnikov:2010vw}
S.~Krasnikov.
\newblock {No to censorship! Comment on the Friedman-Schleich-Witt theorem}.
\newblock {\em Grav. Cosmol.}, 19:54, 2013.

\bibitem{PhysRevD.58.104016}
Masaru Siino.
\newblock Topology of event horizons.
\newblock {\em Phys. Rev. D}, 58:104016, Oct 1998.

\bibitem{Bohn:2016soe}
Andy Bohn, Lawrence~E. Kidder, and Saul~A. Teukolsky.
\newblock {Toroidal Horizons in Binary Black Hole Mergers}.
\newblock {\em Phys. Rev. D}, 94(6):064009, 2016.

\bibitem{Husa_1999}
Sascha Husa and Jeffrey Winicour.
\newblock Asymmetric merger of black holes.
\newblock {\em Phys. Rev. D}, 60:084019, Sep 1999.

\bibitem{BachWeyl}
Rudolf Bach and Hermann Weyl.
\newblock {Neue Lösungen der Einsteinschen Gravitationsgleichungen}.
\newblock {\em Math.Z.}, 13:134, 1922.

\bibitem{1964NCim...33..331I}
W.~{Israel} and K.~A. {Khan}.
\newblock {Collinear particles and bondi dipoles in general relativity}.
\newblock {\em Il Nuovo Cimento}, 33(2):331--344, July 1964.

\bibitem{PhysRevD.35.455}
R.~C. Myers.
\newblock Higher-dimensional black holes in compactified space-times.
\newblock {\em Phys. Rev. D}, 35:455--466, Jan 1987.

\bibitem{KRAMER1980259}
D.~Kramer and G.~Neugebauer.
\newblock The superposition of two kerr solutions.
\newblock {\em Physics Letters A}, 75(4):259--261, 1980.

\bibitem{Wald:1972sz}
Robert~M. Wald.
\newblock {Gravitational spin interaction}.
\newblock {\em Phys. Rev. D}, 6:406--413, 1972.

\bibitem{Herdeiro:2008kq}
Carlos A.~R. Herdeiro and Carmen Rebelo.
\newblock {On the interaction between two Kerr black holes}.
\newblock {\em JHEP}, 10:017, 2008.

\bibitem{Costa:2009wj}
Miguel~S. Costa, Carlos A.~R. Herdeiro, and Carmen Rebelo.
\newblock {Dynamical and Thermodynamical Aspects of Interacting Kerr Black
  Holes}.
\newblock {\em Phys. Rev. D}, 79:123508, 2009.

\bibitem{Hennig:2019knn}
J.~\"org Hennig.
\newblock {On the balance problem for two rotating and charged black holes}.
\newblock {\em Class. Quant. Grav.}, 36(23):235001, 2019.

\bibitem{Vigano:2022hrg}
Adriano Vigan\`o.
\newblock {\em {Black Holes and Solution Generating Techniques}}.
\newblock PhD thesis, Milan U., 2022.

\bibitem{Astorino:2021dju}
Marco Astorino and Adriano Vigano.
\newblock {Binary black hole system at equilibrium}.
\newblock {\em Phys. Lett. B}, 820:136506, 2021.

\bibitem{Annulli:2023ydz}
Lorenzo Annulli and Carlos A.~R. Herdeiro.
\newblock {Non-linear tides and Gauss-Bonnet scalarization}.
\newblock {\em Phys. Lett. B}, 845:138137, 2023.

\bibitem{Majumdar:1947eu}
S.~D. Majumdar.
\newblock {A class of exact solutions of Einstein's field equations}.
\newblock {\em Phys. Rev.}, 72:390--398, 1947.

\bibitem{Papapetrou:1948jw}
A.~Papapetrou.
\newblock {Einstein's theory of gravitation and flat space}.
\newblock {\em Proc. Roy. Irish Acad. A}, 52:11--23, 1948.

\bibitem{Herdeiro:2014goa}
Carlos A.~R. Herdeiro and Eugen Radu.
\newblock {Kerr black holes with scalar hair}.
\newblock {\em Phys. Rev. Lett.}, 112:221101, 2014.

\bibitem{Herdeiro:2023mptSch}
Carlos A.~R. Herdeiro and Eugen Radu.
\newblock Two schwarzschild-like black holes balanced by their scalar hair.
\newblock {\em Phys. Rev. D}, 107:064044, Mar 2023.

\bibitem{Herdeiro:2023roz}
Carlos A.~R. Herdeiro and Eugen Radu.
\newblock {Two Spinning Black Holes Balanced by Their Synchronized Scalar
  Hair}.
\newblock {\em Phys. Rev. Lett.}, 131(12):121401, 2023.

\bibitem{Carter:1970ea}
Brandon Carter.
\newblock {The commutation property of a stationary, axisymmetric system}.
\newblock {\em Commun. Math. Phys.}, 17:233--238, 1970.

\bibitem{Wald:1984rg}
Robert~M. Wald.
\newblock {\em {General Relativity}}.
\newblock Chicago Univ. Pr., Chicago, USA, 1984.

\bibitem{Cunha:2017qtt}
Pedro V.~P. Cunha, Emanuele Berti, and Carlos A.~R. Herdeiro.
\newblock {Light-Ring Stability for Ultracompact Objects}.
\newblock {\em Phys. Rev. Lett.}, 119(25):251102, 2017.

\bibitem{Note1}
Here a distinction is made between degenerate and non-degenerate critical
  points. A critical point is said to be degenerate if the determinant of the
  Hessian vanishes at that point. In this work we neglect the analysis of such
  points since the occurrence of such points requires very specific conditions,
  and make the analysis more nuanced. For example, the existence of such a
  point may allow an ultracompact object to possess a single LR \cite
  {Hod:2017zpi}, corresponding to the coalescence of the two LRs predicted by
  \cite {Cunha:2017qtt}.

\bibitem{Ghosh:2021txu}
Rajes Ghosh and Sudipta Sarkar.
\newblock {Light rings of stationary spacetimes}.
\newblock {\em Phys. Rev. D}, 104(4):044019, 2021.

\bibitem{Tavlayan:2022hzl}
Aydin Tavlayan and Bayram Tekin.
\newblock {Light rings around five dimensional stationary black holes and naked
  singularities}.
\newblock {\em Phys. Rev. D}, 107(2):024016, 2023.

\bibitem{Wei:2022mzv}
Shao-Wen Wei and Yu-Xiao Liu.
\newblock {Topology of equatorial timelike circular orbits around stationary
  black holes}.
\newblock {\em Phys. Rev. D}, 107(6):064006, 2023.

\bibitem{Yin:2023pao}
Jiayu Yin, Jie Jiang, and Ming Zhang.
\newblock {Kinematic topologies of black holes}.
\newblock {\em Phys. Rev. D}, 108(4):044077, 2023.

\bibitem{Wei:2022dzw}
Shao-Wen Wei, Yu-Xiao Liu, and Robert~B. Mann.
\newblock {Black Hole Solutions as Topological Thermodynamic Defects}.
\newblock {\em Phys. Rev. Lett.}, 129(19):191101, 2022.

\bibitem{Note2}
The absence of conical singularities corresponds to requiring that sufficiently
  near the axis the proper perimeter, $\protect \mathcal {P}$, and radius,
  $\protect \mathcal {R}$, of a small circumference of constant $t,r,z$ are
  related by the usual Euclidean formula, $\protect \mathcal {P}=2\pi \protect
  \mathcal {R}$. For the line element (\ref {eq:LineEl}) we have \begin
  {equation} \protect \mathcal {P} = 2\pi \protect \sqrt {g_{\phi \phi
  }}\protect \,,\protect \tmspace +\thickmuskip {.2777em} \protect \mathcal {R}
  = \intop _0^\rho \protect \sqrt {g_{\rho \rho }}\protect \mathrm {d}\rho
  \simeq \rho \protect \sqrt {g_{\rho \rho }|_{\rho =0}} \protect \,. \end
  {equation} Thus, near $\rho =0$ one has $g_{\phi \phi }\left (\rho ,z\right
  )\simeq g_{\rho \rho }\left (0,z\right )\rho ^2+...$.

\bibitem{JaumeCarot_2000}
Jaume Carot.
\newblock Some developments on axial symmetry.
\newblock {\em Classical and Quantum Gravity}, 17(14):2675, jul 2000.

\bibitem{Note3}
Gaussian normal coordinates are a convenient set of coordinates naturally
  adapted to some hypersurface, $\protect \mathcal {S}$, with normal vector
  $k$. To construct these coordinate system we consider at each point $p\in
  \protect \mathcal {S}$ (parameterized by some coordinates $\{u^1,u^2,u^3\}$
  on $\protect \mathcal {S}$) the geodesic whose tangent vector is precisely
  $k$. Then, denoting the affine parameter along these geodesics as $X$, such
  that $X|_{\protect \mathcal {S}}=0$, each point in the neighborhood of
  $\protect \mathcal {S}$ has coordinates $\{u^1,u^2,u^3,X\}$. Eventually these
  coordinate system may become ill defined if the geodesics focus and cross,
  but they will always be valid in some neighborhood of $X=0$. For a deeper
  discussion see~\cite {Wald:1984rg} or~\cite {Carroll:2004st}.

\bibitem{Medved:2004tp}
A.~J.~M. Medved, Damien Martin, and Matt Visser.
\newblock {Dirty black holes: Symmetries at stationary nonstatic horizons}.
\newblock {\em Phys. Rev. D}, 70:024009, 2004.

\bibitem{Carmo}
Manfredo P.~Do Carmo.
\newblock {\em Differential Geometry of Curves \& Surfaces: Revised \& Updated
  Second Edition}.
\newblock Dover Publications Inc., 2016.

\bibitem{Hawking:1973uf}
Stephen~W. Hawking and George F.~R. Ellis.
\newblock {\em {The Large Scale Structure of Space-Time}}.
\newblock Cambridge Monographs on Mathematical Physics. Cambridge University
  Press, 2 2023.

\bibitem{KLEIHAUS2019134892}
Burkhard Kleihaus, Jutta Kunz, and Eugen Radu.
\newblock Balancing a static black ring with a phantom scalar field.
\newblock {\em Physics Letters B}, 797:134892, 2019.

\bibitem{Emparan:2006mm}
Roberto Emparan and Harvey~S. Reall.
\newblock {Black Rings}.
\newblock {\em Class. Quant. Grav.}, 23:R169, 2006.

\bibitem{Note4}
There are surfaces with zero Euler characteristic which are no tori, for
  example the Klein bottle or the M{\"o}bius strip.

\bibitem{Cunha:2022nyw}
Pedro V.~P. Cunha, Carlos A.~R. Herdeiro, and Jo\~ao P.~A. Novo.
\newblock {Null and timelike circular orbits from equivalent 2D metrics}.
\newblock {\em Class. Quant. Grav.}, 39(22):225007, 2022.

\bibitem{Cunha:2016bpi}
Pedro V.~P. Cunha, Carlos A.~R. Herdeiro, Eugen Radu, and Helgi~F. Runarsson.
\newblock {Shadows of Kerr black holes with and without scalar hair}.
\newblock {\em Int. J. Mod. Phys. D}, 25(09):1641021, 2016.

\bibitem{Weyl}
Hermann Weyl.
\newblock Zur gravitationstheorie.
\newblock {\em Annalen der Physik}, 359(18):117--145, 1917.

\bibitem{Emparan:2001wk}
Roberto Emparan and Harvey~S. Reall.
\newblock {Generalized Weyl solutions}.
\newblock {\em Phys. Rev. D}, 65:084025, 2002.

\bibitem{Harmark:2004rm}
Troels Harmark.
\newblock {Stationary and axisymmetric solutions of higher-dimensional general
  relativity}.
\newblock {\em Phys. Rev. D}, 70:124002, 2004.

\bibitem{Note5}
Semi-infinite timelike rods describe an accelerating BH.

\bibitem{Chen:2012dr}
Yu~Chen and Edward Teo.
\newblock {Balanced electric-magnetic dihole in Kaluza-Klein theory}.
\newblock {\em JHEP}, 09:085, 2012.

\bibitem{Hertog:2019hfb}
Thomas Hertog, Tom Lemmens, and Bert Vercnocke.
\newblock {Imaging Higher Dimensional Black Objects}.
\newblock {\em Phys. Rev. D}, 100(4):046011, 2019.

\bibitem{Hod:2017zpi}
Shahar Hod.
\newblock {On the number of light rings in curved spacetimes of ultra-compact
  objects}.
\newblock {\em Phys. Lett. B}, 776:1--4, 2018.

\bibitem{Carroll:2004st}
Sean~M. Carroll.
\newblock {\em {Spacetime and Geometry}: {An Introduction to General
  Relativity}}.
\newblock Cambridge University Press, 7 2019.

\bibitem{garfken67:math}
George Arfken.
\newblock {\em Mathematical Methods for Physicists}.
\newblock Academic Press, {Inc.}, San Diego, third edition, 1985.

\end{thebibliography}
	\pagebreak

	\appendix
	
	\section{Toroidal coordinates\label{sec:TorCoord}}
	
	Toroidal coordinates, $\left\{ \tau,\sigma,\phi\right\} $, are a three-dimensional orthogonal coordinate system~\cite{garfken67:math}, resulting from the rotation of the two-dimensional bipolar coordinate system around the central axis that separates the two focci, $F_{1}$ and $F_{2}$. These focci become a ring of radius $a$ on the $XY$ plane of the Cartesian coordinate system $\{X,Y,Z\}$, where the $Z$ axis is the symmetry axis. The toroidal coordinates lie on the following ranges, $\tau>0\,,-\pi<\sigma<\pi\,,0<\phi<2\pi$. The toroidal coordinates are related to the Cartesian ones via the relations
	\begin{align}
		X & =a\frac{\sinh\tau}{\cosh\tau-\cos\sigma}\cos\phi\nonumber \\
		Y & =a\frac{\sinh\tau}{\cosh\tau-\cos\sigma}\sin\phi\label{eq:TorCart}\\
		Z & =a\frac{\sin\sigma}{\cosh\tau-\cos\sigma}\nonumber 
	\end{align}
	and to cylindrical coordinates via
	\begin{align}
		\rho & =a\frac{\sinh\tau}{\cosh\tau-\cos\sigma}\nonumber \\
		Z & =a\frac{\sin\sigma}{\cosh\tau-\cos\sigma} \label{eq:TorCyl}\\
		\phi & =\phi\nonumber 
	\end{align}
	
	It is helpful to study the coordinate surfaces on this coordinate system. The surfaces of constant $\phi$ correspond to the planes
	\begin{equation}
		Y=X\tan\phi\,.
	\end{equation}
	The coordinates of constant $\sigma$ correspond to spheres of different radii
	\begin{equation}
		\left(X^{2}+Y^{2}\right)+\left(Z-a\cot\sigma\right)^{2}=\frac{a^{2}}{\sin^{2}\sigma}\,,
	\end{equation}
	which all pass through the focal ring, $X^{2}+Y^{2}=a^{2}$, but are not concentric. The surfaces of constant $\tau$ are non-intersecting tori of different radii
	\begin{equation}
		Z^{2}+\left(\rho^{2}-a\coth\tau\right)^{2}=\frac{a^{2}}{\sinh^{2}\tau}\,,
	\end{equation}
	that surround the focal ring. 
	
	In toroidal coordinates the Minkowski spacetime reads:
	\begin{equation}
		{\rm d}s^{2}=-{\rm d}t^{2}+\frac{\sinh^{2}\tau}{\left(\cos\sigma-\cosh\tau\right)^2}{\rm d}\phi^{2}+\frac{{\rm d}\tau^{2}+{\rm d}\sigma^{2}}{\left(\cos\sigma-\cosh\tau\right)^{2}}\,.
	\end{equation}
	It should be noted that spatial infinity in this coordinate system is located at $\tau=0\,,\sigma=0$.

	\section{Brief study of the toroidal BH metric \label{app:Eugen}}
	
	This appendix is devoted to the study of some properties of the metric (\ref{eq:Eugen}), originally proposed in~\cite{KLEIHAUS2019134892} in ring-like coordinate $\{x,y\}$, with the transformation to toroidal-like coordinates being:
	\begin{equation}
		x=-\cos \sigma\,,\quad y=-\cosh \tau\,.
	\end{equation}
	The metric  (\ref{eq:Eugen}) was written using a combination of both ring and toroidal-like coordinates.
 
         Regarding the regularity of the spacetime, the expressions for some curvature invariants, namely Ricci scalar and Kretschmann invariant, were analysed, and we concluded that as long as the $\Lambda$ is smooth (with smooth derivatives as well) and strictly positive function, then the line element (\ref{eq:Eugen}) appears to be regular on and outside the horizon.  The explicit expressions are fairly long and not particularly elucidating, and will not be displayed here.
 
        Below we further aim to discuss: i) the horizon embedding and geometry, ii) the nature of the Killing horizon, iii) the asymptotic behaviour, iv) the BH shadow and lensing images of the spacetime.

	\subsection{The geometry of the Killing horizon}
	
	The line element (\ref{eq:Eugen}) contains a Killing horizon at $y=-1/\lambda$. What is the geometry of this surface?
 By absorbing the scale factor $R$ in a metric redefinition, the spatial cross sections of the horizon are described by the line element:
	\begin{equation}
		{\rm d}s^{2}_\mathcal{H}=\frac{1}{\left(\frac{1}{\lambda}-\cos\sigma\right)^{2}}{\rm d}\sigma^{2}+\frac{1+\lambda}{ \Lambda|_{y=-1/\lambda}}{\rm d}\phi^{2}\,. \label{eq:HorMetric}
	\end{equation}
Since $0<\lambda<1$ and the function $\Lambda$ is strictly positive, the metric (\ref{eq:HorMetric}) has no apparent singularities and is everywhere Euclidean.

We can consider the isometric embedding of this surface in Euclidean 3-space $\mathbb{E}^3$, in order to have further insights on its geometry. Standard methods of differential geometry found in~\cite{Carmo} will be applied in our computations. We start by noting that the line element (\ref{eq:HorMetric}) describes a surface of revolution. Such surfaces are formed by revolving a given curve $\alpha$ around the axis of revolution, the curve $\alpha$ is known as the generating curve. For a generic $\alpha(\sigma)=\left(X(\sigma),Y(\sigma)\right)$, where $\sigma$ parameterizes the curve, a possible surface of revolution is 
	\begin{equation}
		\mathcal{S}\left(\sigma,\phi\right)=\left(X\left(\sigma\right)\cos\phi,X\left(\sigma\right)\sin\phi,Y\left(\sigma\right)\right)
	\end{equation}
	Assuming the surface $\mathcal{S}$ is defined in $\mathbb{E}^3$ the induced metric on it is:
	\begin{equation}
		\mathrm{d}s^2_{\mathrm{ind}}=\left(X^{\prime\,2}+Y^{\prime\,2}\right)\mathrm{d}\sigma^2+X^2\mathrm{d}\phi^2\,.
        \label{eq:SE3}
	\end{equation} 
	By comparing~\eqref{eq:SE3} with~\eqref{eq:HorMetric}, we obtain
	\begin{align}
		X&=\sqrt{\frac{1+\lambda}{ \Lambda|_{y=-1/\lambda}}}\,,\\
		Y^{\prime}&=\sqrt{\frac{1}{\left(\frac{1}{\lambda}-\cos\sigma\right)^{2}}-\frac{\left(1+\lambda\right)}{4  \Lambda|_{y=-1/\lambda}^{3}}\left(\Lambda_{,\sigma}|_{y=-1/\lambda		}\right)^{2}}\,.
	\end{align}
	For the choice of $\Lambda$ used in the main text, $i.e.$ in~\eqref{eq:LambdaChoice}, $Y$ increases monotonically, $Y^\prime>0$. This means that the generating curve $\alpha$ and the corresponding surface of revolution $\mathcal{S}$ are both not closed, see Fig. (\ref{fig:ToroidalEmb}).
	
	\begin{center}
		\begin{figure}
			\begin{centering}
				\includegraphics[width=0.3\textwidth]{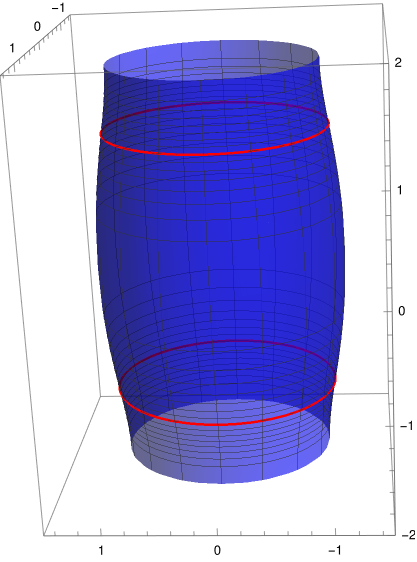}
				\par\end{centering}
			\caption{The isometric embedding of the horizon in (\ref{eq:Eugen}), for $\lambda=1/2\,,R=1\,,M=1$. The red lines indicate the set of points where the Gaussian curvature of the surface vanishes. \label{fig:ToroidalEmb}}
		\end{figure}
		\par
	\end{center}
	
	The embedding at first sight appears to suggest that the horizon surface is not compact. However one can see that $\mathcal{S}$ is indeed a compact surface by recalling that there exists an identification $\sigma\sim\sigma+2\pi$, joining together the points at the top and bottom of $\mathcal{S}$, see Fig. (\ref{fig:ToroidalEmb}).
 
 This apparent problem is not unique to the metric (\ref{eq:HorMetric}). Consider for example the flat torus, whose metric is $\mathrm{d}s^2=\mathrm{d}x^2+\mathrm{d}y^2$, where $\{x,y\}$ are the usual Cartesian coordinates, but with the identification $\left(x,y\right)\sim\left(x+1,y\right)\sim\left(x,y+1\right)$. The isometric embedding of this torus will also not be evidently compact: in fact it will be an open cylinder directed along the symmetry axis. However, similarly as before, an identification of the open-ends has to be considered.  This is also less surprising once one notices that the Gaussian curvature of the flat torus vanishes everywhere. Since the isometric embedding preserves the Gaussian curvature, it is then clear that the embedding of the flat torus must also have vanishing Gaussian curvature everywhere, and so it cannot be an usual toroidal surface in $\mathbb{E}^3$. This statements assume that the embedding is at least $C^2$ smooth.

	\subsection{Nature of the horizon}
	Does the metric~\eqref{eq:Eugen} describe a spacetime with a toroidal BH?
	To assess if that is indeed the case, we shall examine whether the horizon at $y=-1/\lambda$ is compatible with an {\it apparent horizon}, $i.e.$ whether the expansion of outward-pointing, future directed, null geodesic congruences at the horizon's surface vanishes or not.
 
To study such congruences a typical approach is usually to move to horizon-penetrating radial null coordinates. However in the spacetime (\ref{eq:Eugen}) it is hard to construct such a coordinate system, since toroidally-outgoing null curves, $i.e.$ integral lines of the $y$ coordinate, are (in general) not geodesics, and there is no known coordinate chart for which the geodesic equations decouple and allow full integrability. 

Nevertheless, we can overcome this challenge by virtue of a limiting procedure, where we consider the following toroidally-outgoing null curves, which are not necessarily geodesics, in the coordinate system $(t,\sigma,\phi,y)$:
	\begin{align}
		v&=\left(1,0,0,\sqrt{\frac{-g_{tt}}{g_{yy}}}\right)\nonumber\\
		&=\left(1,0,0,\frac{\sqrt{y^2-1} (-y-\cos\sigma) \Delta(y)}{R\left(1-\lambda\cos\sigma\right)}\right)\,.\label{eq:NullVecFie}
	\end{align}
where we have defined $\Delta(y) \equiv 1+ \lambda y$. These curves are future directed and point outwards with respect to closed toroidal surfaces, since $v^y>0$. The parallel transport equation yields
	\begin{align}
		v^\mu\nabla_\mu v^\nu &= \frac{\sqrt{y^2-1} (-y-\cos \sigma) \left(\Delta \Lambda_{,y}-\Delta^\prime \Lambda\right)}{R(1-\lambda  \cos \sigma)
			\Lambda} v^\nu\nonumber\\
		&\quad +\sin \sigma \Delta \frac{(-y-\cos \sigma )}{R(1-\lambda 
			\cos \sigma) \Lambda} \delta^\nu_\sigma\,.
	\end{align}
	At the horizon the geodesic equation is satisfied, and the vector field $v$ is tangent to a congruence of outgoing null geodesics, with 
	\begin{equation}
		\left.v^\mu \nabla_\mu v^\nu\right|_{H}=f_H \left.v^\nu\right|_H\,, \quad f_H=-\sqrt{\frac{1}{\lambda^2}-1}\frac{\Delta^\prime|_H }{R\lambda }\,.
	\end{equation}
	The expansion of this congruence is then 
	\begin{align}
		\Theta&=\nabla_\mu v^\mu -f_H\\
		&=\Delta \frac{2 \Lambda \left(y(y-\cos\sigma)-2\right)+\left(y^2-1\right) (y+\cos \sigma) \Lambda_{,y}}{2 R \sqrt{y^2-1}
			(1-\lambda  \cos\sigma) \Lambda}\,.\nonumber
	\end{align}
	This expression is only valid for geodesics at the horizon, where $\Delta=0$, and thus we have a vanishing null expansion $\Theta=0$ at the Killing horizon, which is indeed consistent with an apparent horizon.
		Interestingly, the surface gravity can also be computed from this procedure as the vector field (\ref{eq:NullVecFie}) reduces to the timelike Killing vector field at the horizon, so $f_H=-2\kappa$ and $\kappa=\sqrt{1-\lambda^2}/2R\lambda$.

	\subsection{Spatial infinity}
	
	As discussed in the main text, spatial infinity corresponds to a singular point in the coordinate system used for the metric~(\ref{eq:Eugen}). In this section we shall expand further on this point.
 
 Furthermore, one should verify that the spatial infinity is topologically spherical, and does not share the same topology as the event horizon. To do so one performs a coordinate transformation to the spherical-like coordinates $\{\tilde{r},\theta,\phi\}$, using the flat space relations:
	\begin{align}
		\cos \sigma & = \frac{r^2-R^2}{\sqrt{\left(r^2-R^2\right)^2+4r^2R^2\cos^2\theta}}\,.\label{eq:thetaofr}\\
		y & = -\frac{r^2+R^2}{\sqrt{\left(r^2-R^2\right)^2+4r^2R^2\cos^2\theta}}\,,\label{eq:yofr}\\
             r &= \tilde{r} + \frac{M}{2}\cos\left(2\theta\right)\,.
	\end{align}
	Doing so, and taking the the asymptotic limit, $\tilde{r}\rightarrow\infty$, the line element becomes

\begin{align}
\mathrm{d}s^2=&-\left(1-\frac{2 M}{\tilde{r}}\right)\mathrm{d}t^2+\left(1 -\frac{2 M \sin^2\theta}{\tilde{r}}\right)\mathrm{d}\tilde{r}^2 \,+ \nonumber \\ 
  & +\mathcal{O}\left(\frac{1}{\tilde{r}}\right)\mathrm{d}\theta\,\mathrm{d}\tilde{r} +\tilde{r}^2\mathrm{d}\theta^2 +\tilde{r}^2\sin^2\theta\mathrm{d}\phi^2\,.
\end{align}
This means that the metric approaches Minkowski spacetime at infinity, and that the Komar mass of the spacetime at spatial infinity is indeed $M$. However, the gravitational curvature decays slower than Schwarzschild.

	\subsection{Shadows of a toroidal BH \label{app:Imaging}}

         In this section we provide further details on how the lensing images of the toroidal BH, displayed in Fig.~\ref{fig:Eugen_Shadows}, are obtained.
         The starting point will be to fix the observer's position and construct a local coordinate system (for details see \emph{e.g.}~\cite{Cunha:2016bpi,Cunha:2018acu}). In order to produce images comparable to previous results in the literature, we set our observer in $r=\mathrm{const.}$ hypersurfaces, where by Eqs. (\ref{eq:thetaofr}) and (\ref{eq:yofr}) the coordinate $r$ is given by:
	\begin{equation}
		r=R\sqrt{\frac{y-\cos\sigma}{y+\cos\sigma}}\,.
	\end{equation}
	The images obtained will be virtually independent on the observer's azimuthal $\phi$ coordinate, due to axial symmetry. However the images will depend on the inclination of the observer with respect to the equatorial plane, determined by the latitude angle $\theta$ according to the flat space relations (\ref{eq:thetaofr}) and (\ref{eq:yofr}), which yield:
	\begin{equation}
		\cos\theta=\frac{\sin\sigma}{\sqrt{y^2-\cos^2\sigma}}\,.
	\end{equation}
	
	The observation frame is defined by a vector basis $\{\hat{e}_{(t)},\hat{e}_{(\phi)},\hat{e}_{(1)},\hat{e}_{(2)}\}$ at the observer's location with a Minkowski normalization:
    \begin{equation}
		\hat{e}_{(\mu)}\cdot\hat{e}_{(\nu)}=\eta_{(\mu)(\nu)}\,,
	\end{equation}
	where $\eta_{(\mu)(\nu)}$ is the Minkowski metric. Both $\hat{e}_{(t)}$ and $\hat{e}_{(\phi)}$ are adapted to the Killing coordinates:
	\begin{equation}
		\hat{e}_{(t)}=\frac{\partial_t}{\sqrt{-g_{tt}}}\,,\quad\hat{e}_{(\phi)}=\frac{\partial_\phi}{\sqrt{g_{\phi\phi}}}\,.
	\end{equation}

In contrast, the vector $\hat{e}_{(1)}$ will be chosen as the ``normal direction'' to $r=\mathrm{const.}$ hypersurfaces, defined via the 1-form:
 \begin{equation}
 \tau = A\,dr =  A \left(r_y dy + r_\sigma d\sigma\right)
 \end{equation}

where $r_y:=\partial r/\partial y$ and $r_\sigma:=\partial r/\partial \sigma$, and whose expressions are:
	\begin{equation}
		\begin{split}
			r_y&=R\frac{ \cos \sigma  }{(\cos\sigma+y)^2}\sqrt{\frac{y+\cos\sigma }{y-\cos \sigma}}\,,\\
			r_\sigma&=R\frac{ y \sin\sigma }{(\cos\sigma+y)^2}\sqrt{\frac{y+\cos\sigma }{y-\cos \sigma}}\,.\\
		\end{split}
	\end{equation}

The vector $\hat{e}_{(1)}$ is then defined via the metric mapping of the 1-form $\tau$ into a vector, $i.e.$ $\hat{e}_{(1)}^\mu = g^{\mu\nu} \tau_\nu$, which yields:
	\begin{equation}
		\hat{e}_{(1)}=A\left( \frac{r_y}{g_{yy}} \partial_y + \frac{r_\sigma}{g_{\sigma\sigma}} \partial_\sigma\right)\,,
	\end{equation}

Further requiring that  $\hat{e}_{(1)}\,dr >0 \implies A>0$, $i.e.$ the vector $\hat{e}_{(1)}$ is pointing outwards, then leads to:
	\begin{equation}
		A = \left(\frac{r_y^2}{g_{yy}}+\frac{r_\sigma^2}{g_{\sigma\sigma}}\right)^{-1/2}\,,
	\end{equation}

The remainder vector $\hat{e}_{(2)}$ is partially fixed by the Minkowski normalization conditions.
 First one writes $\hat{e}_{(2)}=\zeta\,\partial_y + \chi\,\partial_\sigma$, and makes the choice $\chi<0$. Then solving for $\chi$ and $\zeta$ yields:
	\begin{equation}
		\zeta=-\frac{r_\sigma \chi}{r_y}\,,\quad\chi=\frac{-|r_y|}{\sqrt{g_{yy} r^2_\sigma + g_{\sigma\sigma}r^2_y }}\,.
	\end{equation}
	The components of the locally measured 4-momentum, $p_{(a)}=\hat{e}^\mu_{(a)}p_\mu$, are:
	\begin{align}
		p_{(t)}&=\frac{E}{\sqrt{-g_{tt}}}\,,\nonumber\\
		p_{(\phi)}&=\frac{L}{\sqrt{g_{\phi\phi}}}\,,\nonumber\\
		p_{(1)}&=A\left(p_y\frac{ r_y}{g_{yy}}+p_\sigma\frac{ r_\sigma}{g_{\sigma\sigma}}\right)\,,\\
		p_{(2)}&=\zeta\,p_y + \chi\,p_\sigma\,.\nonumber
	\end{align}
	The quantities $E$ and $L$ above represent (respectively) the conserved energy and angular momentum of the photon along each light ray.
	Given the relations above, we can then connect the locally measure 4-momentum in terms of observation angles $\alpha,\beta$ defined within the frame:
	\begin{align}
		p_{(t)}&=\mathrm{p}\,,\nonumber\\
		p_{(\phi)}&=\mathrm{p}\cos\alpha\sin\beta\,,\nonumber\\
		p_{(1)}&=\mathrm{p}\cos\alpha\cos\beta\,,\\
		p_{(2)}&=\mathrm{p}\sin\alpha\,.\nonumber
	\end{align}
		For the purposes of numerically integrating the null geodesics, it is useful to express the 4-velocity components $\dot{x}^\mu=g^{\mu\nu}p_\nu$ in terms of the locally measured 4-momentum. The non-trivial components yield:
	\begin{align}
		\dot{y} & = \frac{\mathrm{p}}{\sqrt{g_{\sigma\sigma} r_y ^2 + g_{yy}r_\sigma^2}}\left(r_\sigma\,\textrm{sign}(r_y)\,\sin\alpha + r_y \sqrt{\frac{g_{\sigma\sigma}}{g_{yy}}}\cos\alpha\cos\beta\right)\,,\nonumber\\
		\dot{\sigma} & = \frac{\mathrm{p}}{\sqrt{g_{\sigma\sigma} r_y ^2 + g_{yy}r_\sigma^2}}\left( -|r_y|\, \sin\alpha + r_\sigma \sqrt{\frac{g_{yy}}{g_{\sigma\sigma}}}\cos\alpha\cos\beta\right)\,.
	\end{align}
	These equations can be propagated backwards in time from the observer's position in order to obtain the lensing image and shadow of the BH.
        We display in Fig.~\ref{fig:Eugen_MoreShadows} some additional images with different inclinations $\theta$.
	
	\onecolumngrid
	\begin{center}
		\begin{figure}
			\begin{centering}
				\includegraphics[width=0.3\textwidth]{Figures/ToroidalBH_th0.png}
				\includegraphics[width=0.3\textwidth]{Figures/ToroidalBH_th0-shadow.png}
				
				\vspace{3pt}
				
				\includegraphics[width=0.3\textwidth]{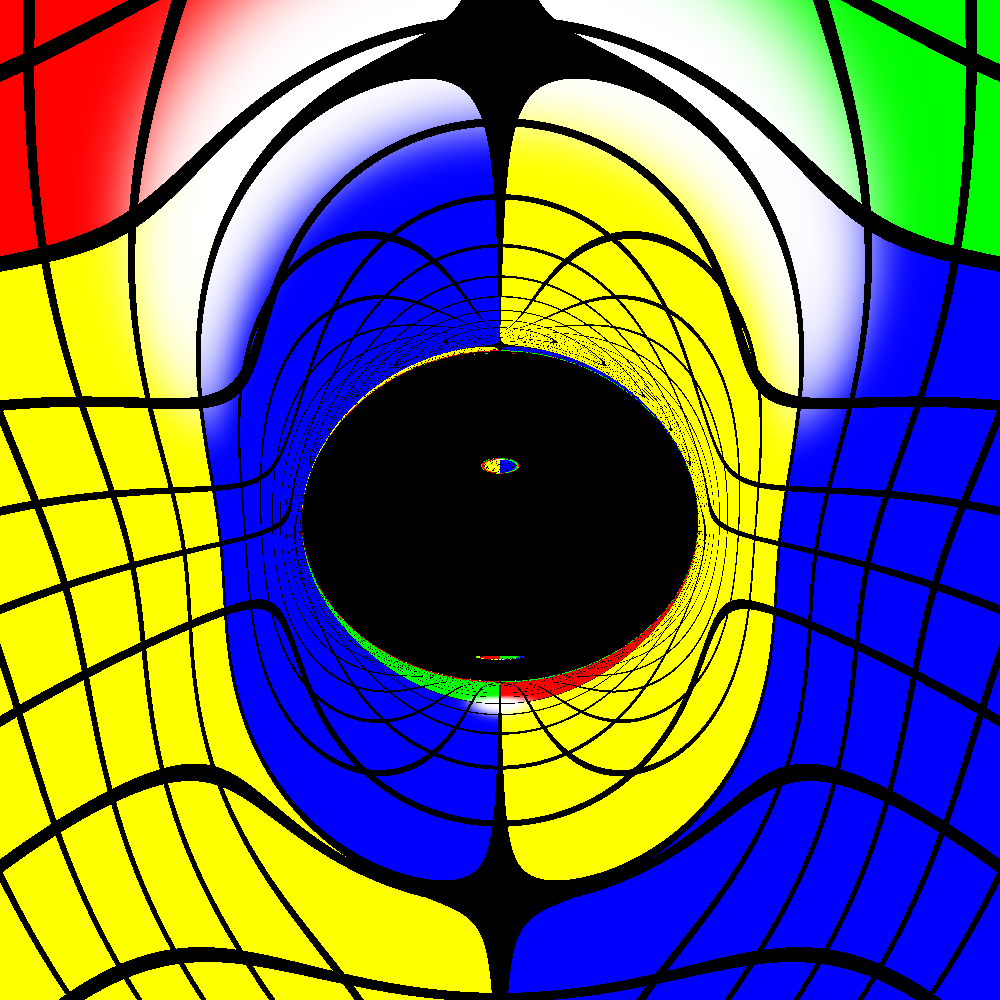}
				\includegraphics[width=0.3\textwidth]{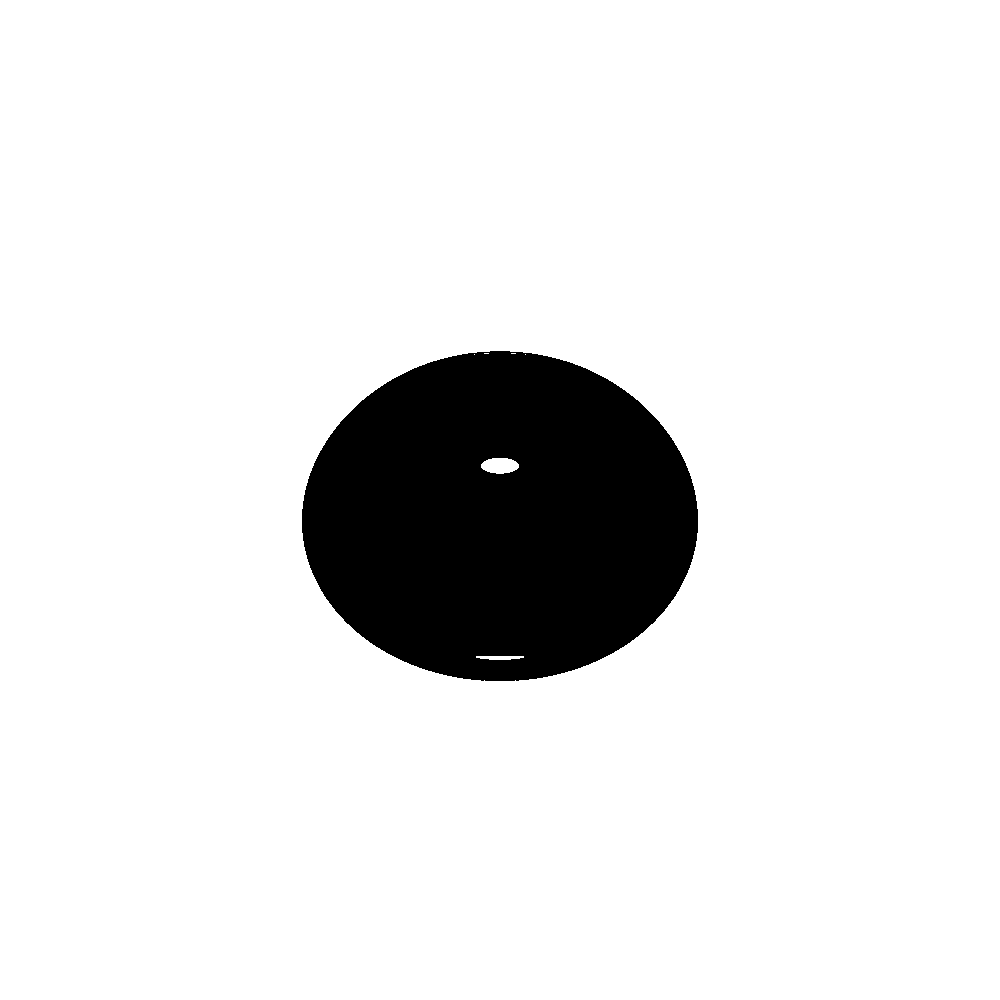}
				
				\vspace{3pt}
				
				\includegraphics[width=0.3\textwidth]{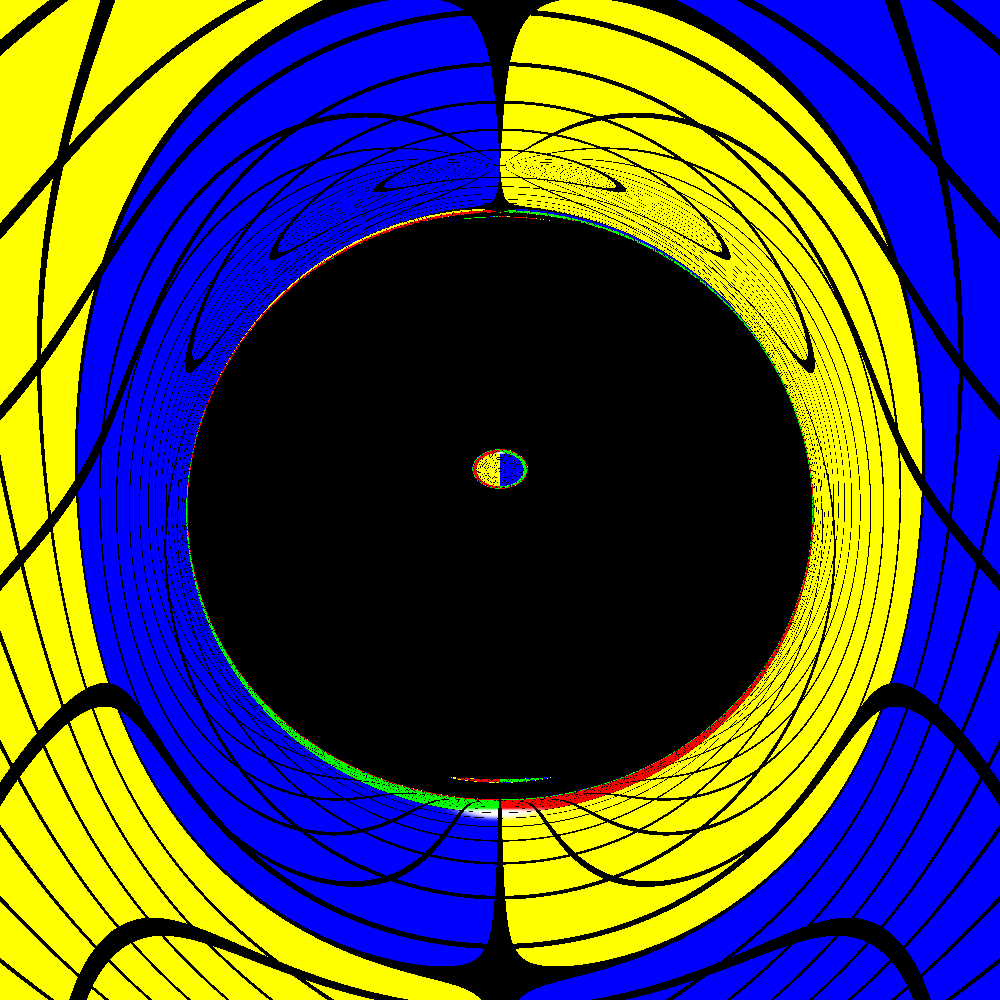}
				\includegraphics[width=0.3\textwidth]{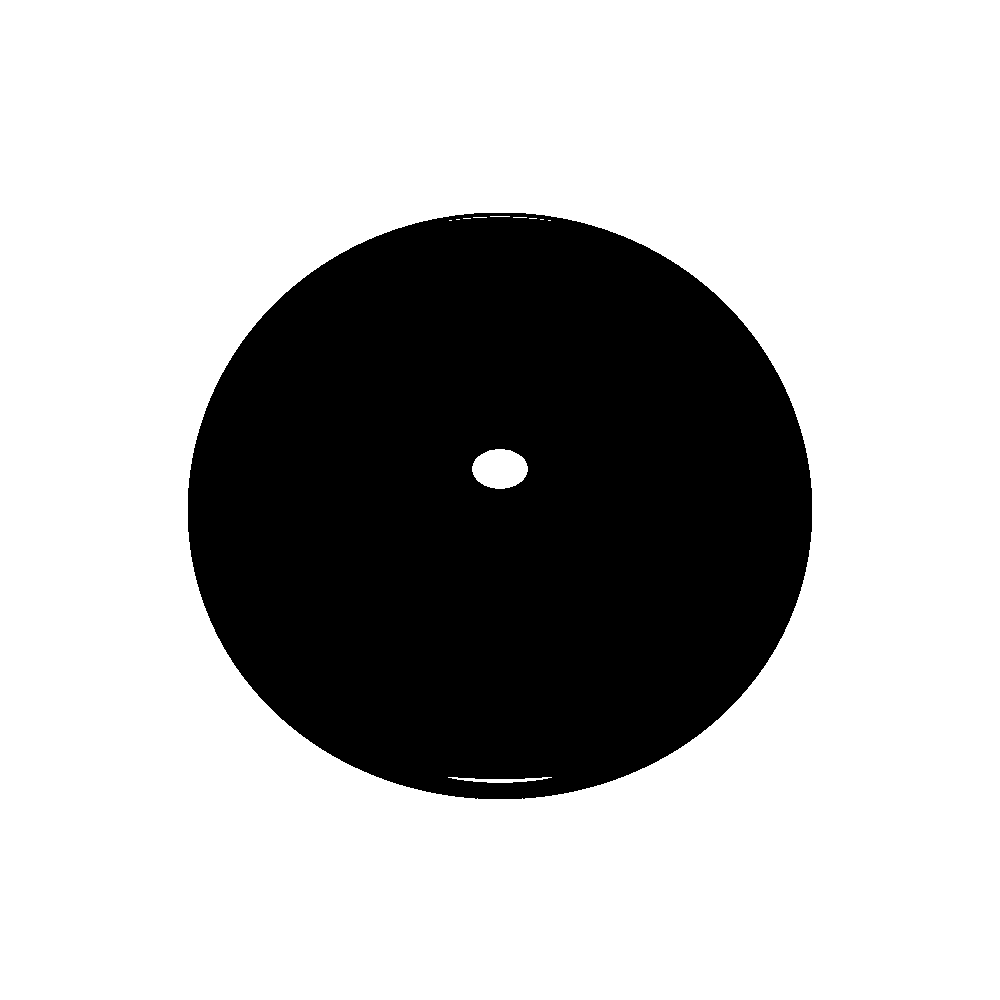}
				
				\vspace{3pt}
				
				\includegraphics[width=0.3\textwidth]{Figures/ToroidalBH_th10.png}
				\includegraphics[width=0.3\textwidth]{Figures/ToroidalBH_th10-shadow.png}
				
				\vspace{3pt}
				\par\end{centering}
			\caption{Shadows of the BHs described by Eq. (\ref{eq:Eugen}), with $\lambda=1/2\,,R=1\,,M=1$, for the inclinations $\theta=\{0,3,7,10\}$ degrees (from top to bottom). The images on the left have a colored celestial sphere, while the ones on the right depict only the shadows.
				\label{fig:Eugen_MoreShadows}}
		\end{figure}
		\par
	\end{center}
	\twocolumngrid
	
\end{document}